\documentclass[prl,twocolumn]{revtex4-2}

\usepackage{epsfig}
\usepackage{graphicx}
\usepackage{amsfonts,amssymb}
\usepackage{amsmath}
\usepackage{color}
\usepackage{times}
\usepackage{flushend}
\usepackage{xcolor}
\usepackage{hyperref}

\hypersetup{
    colorlinks=true, 
    linkcolor=blue, 
    citecolor=blue, 
    filecolor=magenta, 
    urlcolor=blue, 
}

\begin{document}

\title{Enormous enhancement of resistivity in nanostructured 
electron-phonon systems}

\author{Debraj Bose, Sankha Subhra Bakshi  and Pinaki Majumdar}

\affiliation{Harish-Chandra Research Institute
(A CI of Homi Bhabha National Institute), 
Chhatnag Road, Jhusi, Allahabad 211019
}
\date{\today}

\begin{abstract}
Recent experiments on nanoclusters of silver (Ag) embedded in a gold (Au) 
matrix reveal a huge increase in both the zero temperature resistivity and 
the coefficient of the ``$T$ linear'' thermal resistivity with increasing  
volume fraction of Ag. A fraction $f \sim 50\%$ of Ag leads to a factor  
of $20$ increase  in the residual resistivity, and a  $40$ fold enhancement 
in the coefficient of linear $T$ resistivity, with respect to Au. 
Since Au and Ag both have weak electron-phonon coupling we surmise 
that the huge enhancements arise from a moderately large electron-phonon 
coupling that may emerge at the Ag-Au interface. We construct nanocluster 
configurations for varying  $f$ in two dimensions, define a Holstein model 
on it with weak coupling on the `interior' sites and a strong coupling on the 
interfacial sites, and solve the model through exact diagonalisation based
Langevin dynamics. Computing the resistivity, we observe a large $T=0$ increase 
with $f$ and also a linear $T$ enhancement factor of $\sim 30$. While the 
enhancement factors are parameter choice dependent, our key qualitative 
result is that the interface physics is inhomogeneous, with widely varying
distortions, and different segments of the interface dictate the residual 
resistivity and the thermal scattering.  
\end{abstract}

\maketitle


In the noble metals the resistivity around room temperature 
is dominated by electron-phonon scattering
\cite{ziman,allen-transp}. The temperature 
dependence of resistivity
takes the form $\rho(T) \sim \rho_0 + \delta \rho(T)$ where 
the residual resistivity $\rho_0$ in high purity samples is
$ \lesssim 0.1\mu \Omega$cm \cite{res-gold}, 
and the temperature dependence is $\delta \rho(T) \sim 
g^2 T$, where $g$ is the effective electron-phonon (EP) 
coupling \cite{ep-note}. In these materials $g$ is small
\cite{allen-epc}, and  $\delta \rho(300K) \sim$
few $\mu \Omega$cm \cite{res-gold}.
One may expect that nanohybrids 
of two such metals will have a 
similar $\delta \rho(T)$. 
Instead, recent experiments \cite{arind1,arind2}
on clusters of Ag embedded in Au show that in a
hybrid with $50-50$ Ag-Au the linear $T$ coefficient 
is $40$ fold larger than that in Au nanoclusters!

`Anomalous' transport in electron-phonon systems has
been known before, due to strong coupling or the 
presence of disorder. 
The simplest instance is deviation from linear 
$T$ dependence of resistivity, with a reducing
slope suggestive of resistivity saturation
in strong coupling systems
\cite{saturation-allen, saturation-millis,
saturation-gun1,saturation-gun2}. 
The next is polaron formation 
\cite{pol-alexand, pol-franchini, pol-romero, pol-bonca,
pol-ciuchi, pol-millis, pol-pm}
and the emergence of a charge localised
insulating phase. Finally, there are instances of 
disorder interplaying with strong EP coupling
leading to  deviations from Mathiessen's rule,
e.g, in A-15 compounds 
\cite{ep-dis-a15-1, ep-dis-a15-2}.
Several attempts have been made to study the 
disordered strong coupling problem theoretically
\cite{ep-dis-emin,ep-dis-pm,ep-dis-ciuchi}.

The Au-Ag nanohybrids 
share some similarities with the situations above
but also major differences: (i)~at large volume
fraction of Ag in Au $(\sim 50-50)$, one sees
a sublinear $T$ dependence of $\delta \rho(T)$
- the precursor of resistivity saturation.
This is a `strong coupling' signature. However,
(ii)~both Ag and Au are 
wide band metals with weak EP coupling and just 
a hybrid structure would not increase the EP 
coupling in the interior of the clusters.  
Any enhancement would be related to the interfaces.
This situation we feel has both ingredients of 
(a)~emergent strong coupling, and 
(b)~disorder due to the presence of interfaces.
In addition to transport measurements, the
experiments confirmed the enhancement of
effective EP coupling through point contact
spectroscopy \cite{arind2}.

To model this nanohybrid system we
envisage three kinds of sites.
Two of these are `interior sites':  Au and Ag sites
away from the Au-Ag interface, while the third are
Au or Ag sites at the interface. We assume
that the interior sites would
have electronic parameters mimicking bulk Au or Ag, while
the interfacial sites would have different parameters 
altogether.  On a given structural motif a fraction 
$f$ would be Ag and $1-f$ would be Au. 
Of the total number of sites a
certain fraction $I_f$  
would be at the interface. 
Within a two dimensional (2D) Holstein model
we assume that (i)~the onsite potential of 
Au and Ag sites are equal (we change this later), 
(ii)~all `interior' sites (Ag or
Au) have weak EP coupling $g_1$, and 
(iii)~Ag or Au sites at the interface have a 
larger EP coupling $g_2$.
We work at density $n=0.25$ and use $g_1 = 0.2$ and $g_2=1.6$.  
The basis for choosing these numbers, and the effect of 
parameter variations is discussed later. 

We generate structural motifs for varying volume fraction 
$f$ via a simple algorithm described in the Supplement.
We solve for the phonon distortions on these structures 
by using a Langevin equation 
\cite{lang-moz,lang-lu,lang-pm1,lang-pm2}
that iteratively diagonalises the
electron problem and exactly handles strong EP coupling 
and thermal fluctuations.
The distortion fields $x_i^{\alpha}(t)$, where ${\bf R}_i$ is
the site index, $t$ is time, and $\alpha$ denotes an interface
geometry, serve as the input to calculating electronic
properties like resistivity and density of states.

Before studying the nanohybrid we computed
$\rho(T)$ in the `clean' problem for varying EP coupling
- all the way from the weak coupling metal to the polaronic
insulator. The results, and the `scaling' that emerges
for $\rho(T)$, serve as a template for analysing 
the clustered system. 
Our main results on the nanostructured problem 
are the following:

(i)~{\it Nature of the interface:}
Despite the same strong EP coupling on all interfacial 
sites the interface is very
imhomogeneous, with wide variation in local density 
and associated lattice distortions. 

(ii)~{\it Residual resistivity:}
The zero temperature resistivity $\rho_0$ arises from 
large distortions at a small fraction 
of the interfacial sites, which acts
as effective `disorder' seen by the electrons. 
$\rho_0$ does not increase linearly with $I_f$,
particularly at large $f$.

(iii)~{\it $T$ linear coefficient and saturation:}
The coefficient of $T$ linear resistivity increases by a factor
of $30$ from the `pure Au' limit to that of $50-50$ Ag-Au. 
The experimental enhancement at $50-50$ is $\sim 40$
\cite{arind2}.
$\delta \rho(T,f)$ plotted with respect to $A(f)T$ shows 
approximate collapse to a single sublinear curve, suggestive
of resistivity saturation at higher $T$, particularly at
large $f$.

(iv)~{\it Thermal scattering mechanisms:}
The thermal scattering occurs due to lattice fluctuations
{\it around the $T=0$ distortions.} We find that
at a given $T$ low and
high electron density sites on the interface have 
relatively small fluctuation while 
sites with intermediate density, $n_i \sim 0.6$,
have far larger fluctuation. The $T=0$ and finite $T$
scattering sources are complementary.
 
{\it Model and method:}
We study the 2D Holstein model on cluster 
motifs, defining a variable $\eta_i$
such that for Au or Ag sites on the interface $\eta=1$, while for
sites away from the interface (i.e in the bulk) $\eta=0$.
Bulk sites have EP coupling $g_1$ while interfacial sites have 
EP coupling $g_2$.
We have $ H  = H_k + H_{ph} + H_{ep} $, with
\begin{eqnarray}
H & = & -t\sum_{\langle ij \rangle} c^{\dagger}_i c_j
+ \sum_i \big( { {p_i^2} \over {2M}} + {{Kx_i^2} \over 2} \big)
\cr
~~~~~~ && - g_1 \sum_i (1 - \eta_i) n_i x_i - g_2 \sum_i \eta_i n_i x_i
\end{eqnarray}
We set $t=1$ as the reference scale, $M=25$, 
and $K=1$.  We comment later on the effect 
of a site potential difference between Au and Ag. 
The cluster configuration generating algorithm
is described in the Supplement.
Typical non overlapping clusters are $4 \times 4$. 
At large coverage they begin to overlap.

To generate the equilibrium phonon configurations of $H$ 
at some $T$ on these motifs we use a Langevin equation, with 
thermal noise and a dissipation coefficient which satisfy the 
fluctuation-dissipation relation.  
Defining $H_{el} = H_k + H_{ep}$,
\begin{eqnarray}
M {{d^2x_i} \over {dt^2}} 
& = &  - K x_i(t) - { {\partial \langle H_{el} \rangle} 
\over {\partial x_i}}  - M\gamma {{ d x_i} \over {dt}}
+ \xi_i(t) \cr
{ {\partial \langle H_{el} \rangle} \over {\partial x_i}} &=& 
-(g_1 (1- \eta_i) + g_2 \eta_i) \langle n_i \rangle \cr
\cr
\langle n_i \rangle &=& 
\sum_m |\psi_{im}|^2 n_f(\epsilon_m)
\cr
\langle \xi_i(t) \rangle & =& 0,
~~~~\langle \xi_i(t) \xi_j(t') \rangle
= 2 \gamma k_BT \delta_{ij} \delta(t - t')
\end{eqnarray}
We set the damping factor $\gamma=0.2$.
The $\psi_{im}$ are site amplitudes
of the instantaneous
eigenvectors of $H_{el}$ for a given $x_{i}(t)$ and
$\epsilon_m$ are the corresponding eigenvalues.
The Fermi factors $n_f(\epsilon_m)$ are used to
calculate the occupation of the eigenstates.
The non trivial part is calculating 
$ \langle n_i \rangle$ since it requires
diagonalising $H_{el}$ for every move.  This is 
prohibitive for large systems. We benchmark and use 
a scheme where $ \langle n_i \rangle$ is computed by 
diagonalising $H_{el}$
on an update cluster defined around the target site
${\bf R}_i$. This update cluster includes ${\bf R}_i$,
it's 4 nearest neighbours (NN), the 4 NNN, 
and the 4 axial sites beyond that. This cluster 
geometry, and benchmarks of the cluster scheme with
respect to full diagonalisation, 
are shown in the Supplement.
The scheme allows us access to system 
size $30 \times 30$.

\begin{figure}[b]
\centerline{
\includegraphics[width=8cm,height=5.6cm]{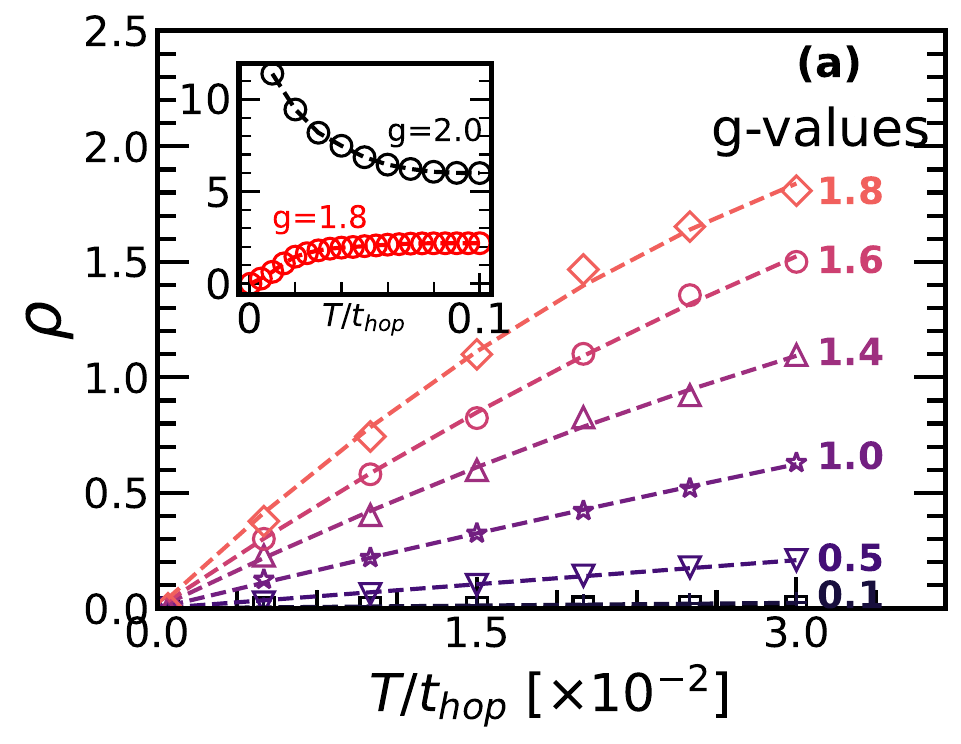}
}
\centerline{
~~~~~
\includegraphics[width=3.2cm,height=4.2cm]{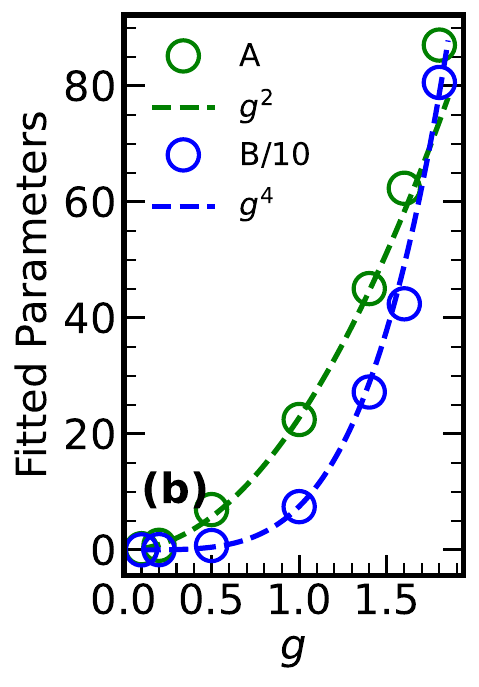}
\includegraphics[width=4.5cm,height=4.2cm]{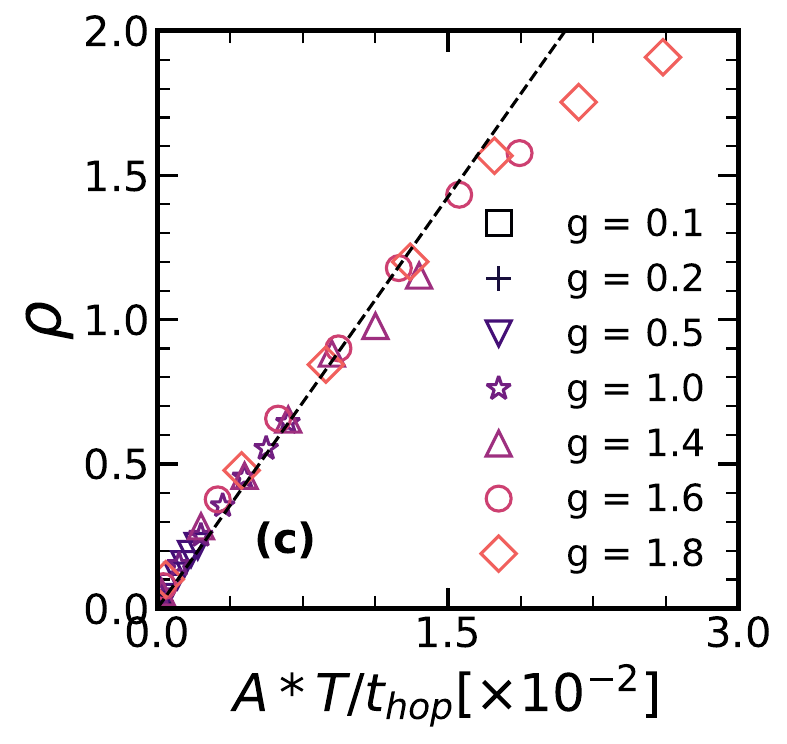}
}
\caption{Resistivity in the homogeneous problem for varying 
EP coupling $g$, at $n=0.25$.
(a)~$\rho(T,g)$, with the dotted lines being fits of the form
$A(g)T - B(g)T^2$. (b)~The coefficients $A(g)$ and $B(g)$ 
fitted to power laws $g^2$ and $g^4$, respectively.
(c)~Plot of $\rho(T,g)$ with respect to $A(g) T$.
There is no perfect scaling collapse, but a rough 
sublinear trend.
}
\end{figure}
%

Fig.1 shows our results on the resistivity $\rho(T)$ in 
the homogeneous problem (all $\eta=0$) for varying 
$g_1=g$. The Kubo formula for conductivity that we use
is given in the Supplement. Our two dimensional 
`resistivity' are in units of $\hbar/(\pi e^2)$.
The main panel in Fig.1(a) shows $\rho(T,g)$, upto $g=1.8$
and $T=0.03$. Over a low $T$ window $\rho(T)$ has a linear
dependence on $T$ in this classical phonon model, unlike
the quantum problem where for $T \ll \theta_D$, the Debye
temperature, it has a faster fall.
With increasing $T$ the resistivity shows a sublinear
behaviour, which gets prominent with increasing $g$.
The inset compares $\rho(T)$ in the metal at $g=1.8$ 
and in the polaronic insulator at $g=2$.
We clearly see the $d\rho/dT \rightarrow 0$ behaviour
at high $T$ in the metal as well as the insulator.
The presence of an insulating state shows why the
metallic resistivity cannot keep indefinitely 
increasing with $T$ and
in this case would saturate to $\rho \sim 4$.

We tried a fit 
$\rho(T) = A(g) T - B(g) T^2$, over the $T=0-0.03t$ window.
The $B$ coefficient reduces the rate of growth so we have
used a negative sign.
The dotted lines in Fig.1(a) show the fit of this form
to $\rho(T,g)$. Fig.1(b) shows $A(g)$ and $B(g)$.  
Perturbation theory suggests that $A(g) \propto g^2$ and
that the next order correction would be $B(g) \propto g^4$.  
The dotted lines are reasonable fits using these power 
laws. 
Fig.1(c) shows $\rho(T,g)$ plotted with respect to 
$A(g)T$. Due to the differing $g$ dependence
of the $A$ and $B$ coefficients there is no `scaling
collapse', but instead a $g$ dependent downward bend.
However, upto $g=1.6$, and for the low $T$ probed,
there is a rough sublinear collapse.


\begin{figure}[b]
\centerline{
\includegraphics[width=8.4cm,height=8.2cm]{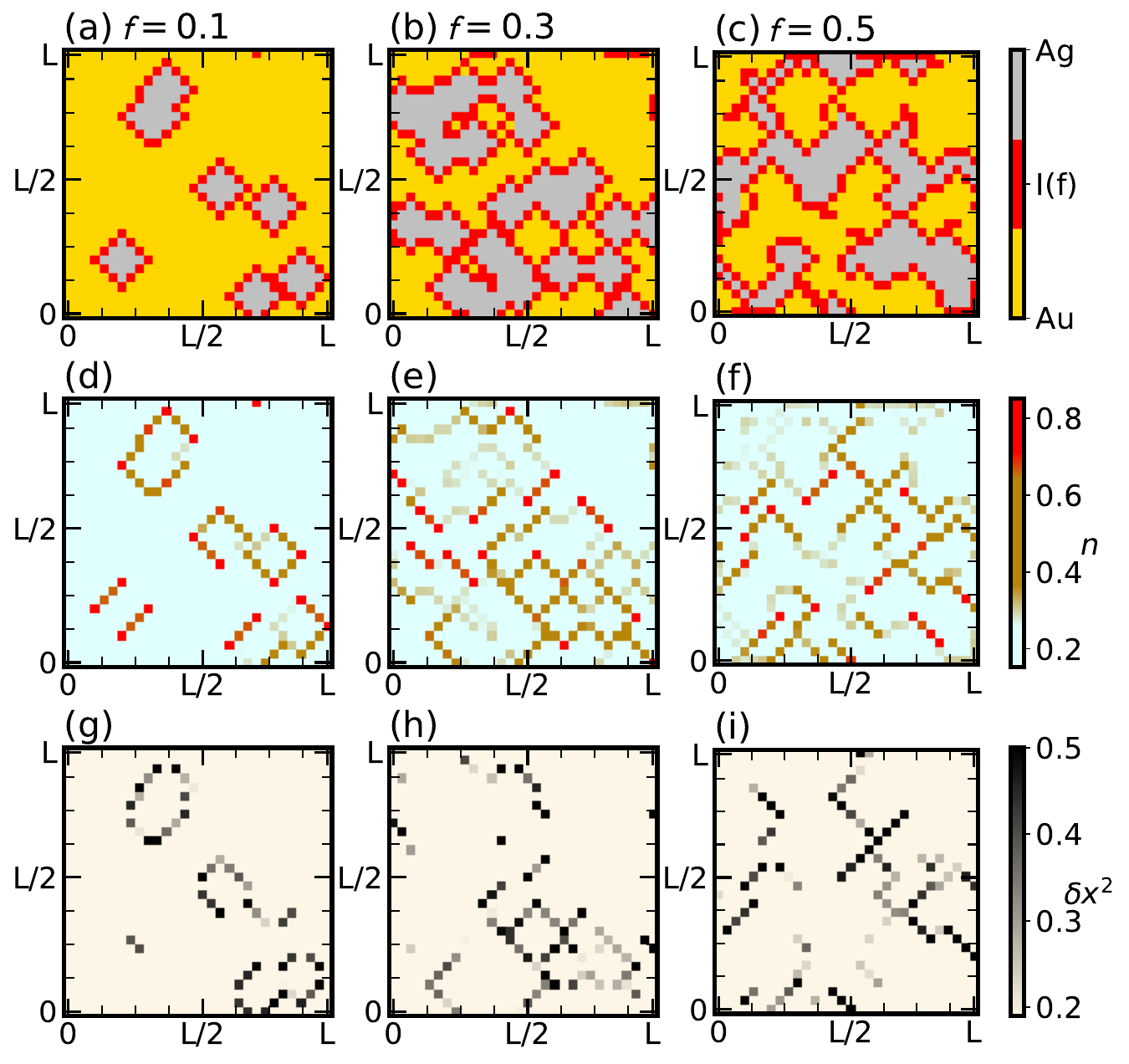}
}
\caption{Interfaces, and spatial character of
distortions at $f=0.1, 0.3, 0.5$.
Top: the `boundary sites' with EP coupling $g_2$.
Middle: Map of density $n_i$ at $T=0$.
Bottom: Map of mean square fluctuation 
$(\Delta x_i)^2 = 
\langle (\delta x_i)^2 \rangle$ at temperature
$T=0.01t$.
Note that all large $g$ locations do not have large
$n_i$, and that large $n_i$ and large 
$(\Delta x_i)^2$ are actually anticorrelated.
}
\end{figure}
\begin{figure}[t]
\centerline{
\includegraphics[width=8.2cm,height=11.0cm]{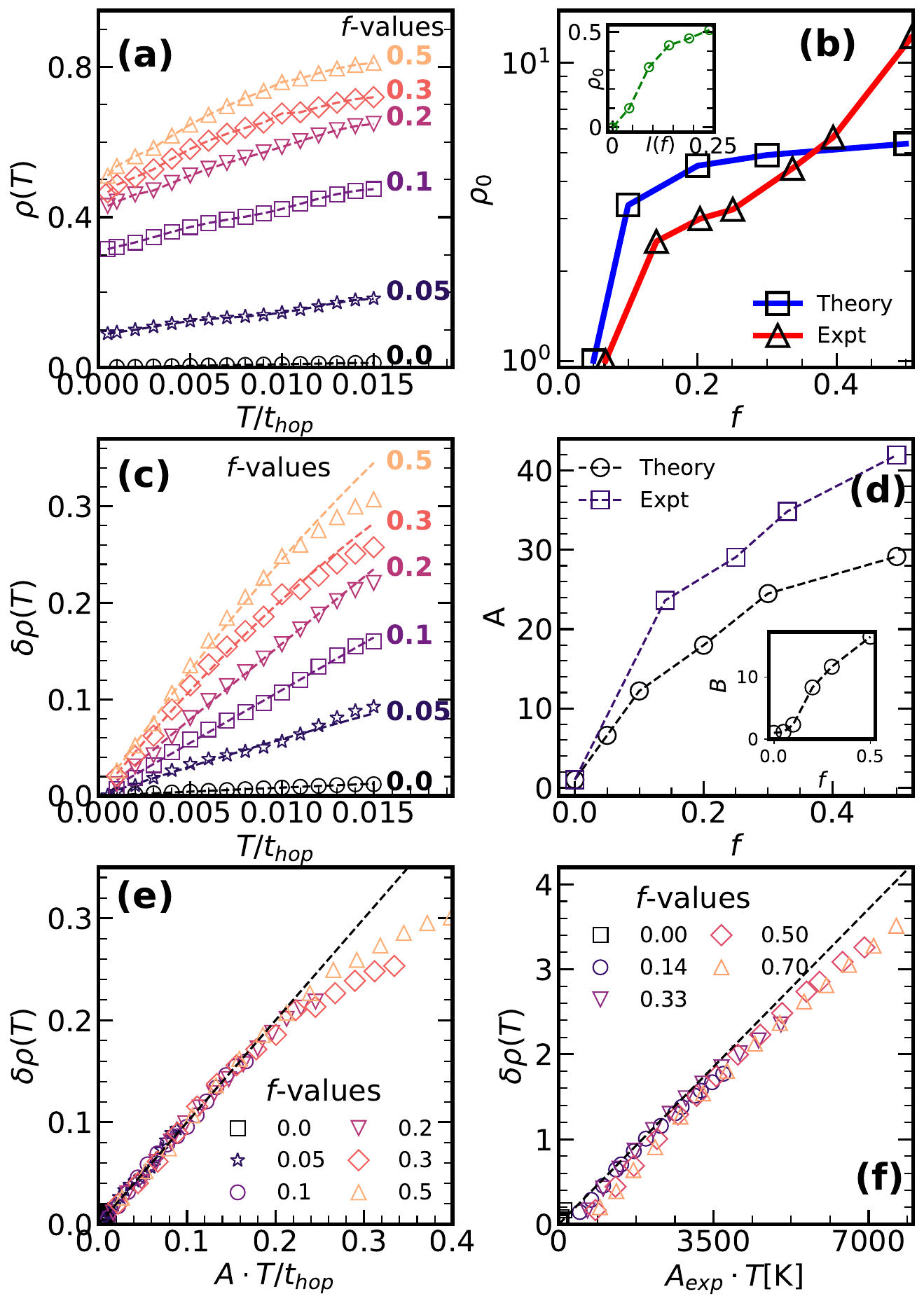}
}
\caption{Resistivity with varying Ag concentration, $f$. 
(a)~The total resistivity $\rho(T,f) = \rho_0(f) + \delta
\rho(T,f)$, on $30 \times 30$ lattices, averaged over 10
cluster configurations. (b)~The residual resistivity, normalised by
it's value at $f=0.05$, comparing our result with experimental
data. The inset shows that our residual resistivity is 
not simply proportional to the perimeter $I_f$. 
(c)~Our result on $\delta \rho(T)$
for  different $f$, note the sublinear growth at
large $f$. We fit $\delta \rho(T,f) = A(f) T - B(f) T^2$ 
(dotted lines).
(d)~$A(f)$, the slope of the linear $T$ part of our 
$\delta \rho(T,f)$,
normalised by it's value at $f=0$, and the corresponding
result from experiment. The inset shows our $B(f)$.
(e)~Plot of our $\delta \rho(T,f)$ with respect 
to $A(f)T$. 
(f)~The experimental result corresponding to panel (e).
}
\end{figure}
%

{\it Cluster configuration and distortions:}
The top row in Fig.2 shows the interface in single
copies of samples
with $f=0.1,~0.3,~0.5$, marked out in red (the large $g$ sites)
and the interior of the Ag clusters marked in silver.
The middle row shows the electron 
density $n_i$ at $T=0$ in these
configurations. While all large $n_i$ sites are on the
interface, not all interfacial sites have large $n_i$.
In fact the density at interfacial sites ranges from
$n \sim 0.2$ (as in the `bulk) all the way to $n \sim 0.8$.
Had the large $g$ sites been {\it isolated} they would all
have had large $n_i$. 
This $n_i$, and the associated $T=0$ distortion field 
$x_i^0$, will be the source of scattering and residual
resistivity.

The bottom row shows the spatially resolved
amplitude of
mean square thermal fluctuation $(\Delta x_i)^2 =
\langle (\delta x_i)^2 \rangle$ at $T=0.01t$,
where $\delta x_i = x_i - x_i^0$.
Again, the large $\Delta x_i$ sites are all on
the interface, but these are only a fraction of
interfacial sites. Further, sites with large (or small)
$n_i$ actually have much smaller $\Delta x_i$ than
sites with $n_i \sim 0.6$.
The thermal scattering, that
leads to the enhanced $T$ linear slope, 
will be dominated by the medium $x_i^0$  
sites on the interface.
All these puts paid to the expectation that 
an interface of length $I_f$ and large 
EP coupling $g_2$ would lead to thermal
scattering $\propto g^2_2 I_f$.
We look next at the resistivity and then
try to relate it to $x_i^0$ and $\Delta x_i$.

{\it Resistivity:}
Fig.3(a) show the resistivity $\rho(T,f)$ that we obtain by
generating $x_i$ configurations on a $30 \times 30$ lattice
and averaging the conductivity 
thermally and over 10 cluster configurations. 
The data is plotted to $T = 0.015t$ 
and shows both the increase in $\rho_0$ with $f$ and
the less prominent increase in the slope of $\delta \rho(T)$.
Fig.3(b) superposes our $\rho_0(f)$ with that obtained
from experiments, both results being normalised to the
value at $f=0.05$. Beyond $f \sim 0.3$ the experimental
$\rho_0$ has a stronger growth than the theory result.
Note that our $f=0$ result has zero resistivity, unlike
the experiment. Our $\rho_0(f)$ does not
vary linearly with $I_f$ at large $f$ (inset).

Fig.3(c) shows the thermal component of $\rho(T)$ 
that we obtain. 
While the coupling constants $g_1$ and $g_2$ are
now fixed we can try a fit $\delta \rho(T,f)
= A(f) T - B(f) T^2$.
The corresponding fits are superposed on the
data points 
Like in the `clean problem' in Fig.1 we see a slope
increasing with $f$ and also sublinear behaviour
at large $T$. 
Fig.3(d) shows the coefficient $A(f)$, normalised by $A(0)$,
comparing our result with experiments. The amplification
factor at $f = 0.5$ is $\sim 40$ experimentally, while for
us it is $\sim 30$. 
This number depends on the choice of $g_2$, as we
comment on in the Supplement. 
The behaviour of our $B(f)$ is shown in the inset to
Fig.3(d).

Fig.3(e) plots our $\delta \rho(T,f)$ with respect 
to $A(f)T$. It shows the deviation from a linear
scaling curve at large $f$ and high $T$. Fig.3(f) 
shows the corresponding scaling curve from
experiments. 

\begin{figure}[b]
\centerline{
\includegraphics[width=8.1cm,height=7.7cm]{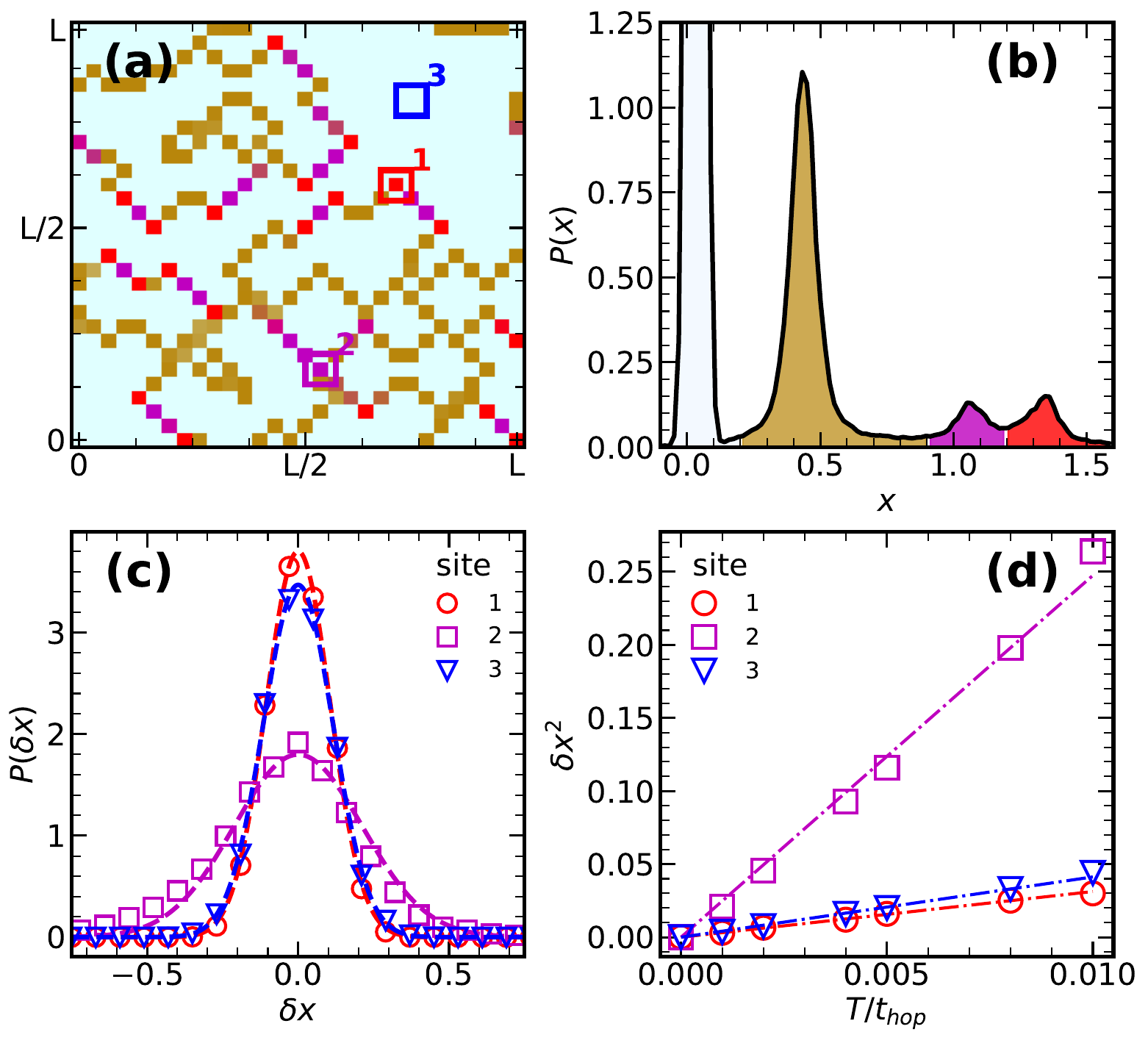}
}
\caption{Phonon distortions and local spectra at $f=0.3$, 
single copy data.
(a)~Map of the $T=0$ distortion $x_i^0$.
(b)~The distribution $P(x)$ of distortions  at $T=0$.
(c)~The local distributions $P_i(\delta x)$, where
$\delta x_i = x_i - x_i^0$ at 3 sites (marked in (a)) 
at $T=0.01t$. (d)~The $T$ dependence of the on site
variance $(\Delta x_i)^2$ at the three sites marked out
in panel (a). All of them are $\propto T$, but the 
slope in the medium density site is much greater
than that in the large and small density sites.
}
\end{figure}

{\it Relation between structural distortions and 
resistivity:}
We take a look at the ditribution of
structural distortions both at zero and finite $T$
and then attempt to relate the distortions to the
resistivity.
Fig.4 looks at the phonon properties
in a particular cluster configuration at $f=0.3$.
Panel (a) looks at the spatial map of
$T=0$ distortions $x_i^0$. 
We use a `four color' map,
identifying distortions as large (red), medium 
(magenta), small (gold) or very small (white). 
Note that at $T=0$ the distortions and local
density are related by $x_i^0 = (g_i/K) n_i$,
obtained simply by miminimising $H_{el}$ with 
respect to $x_i^0$.
 
The distribution of $T=0$ distortion:
$P(x) = (1/N)\sum_i \delta(x - x_i^0)$ is shown
in panel (b). The continuous values of $x_i^0$ are
retained but the features are coloured according to
our large-medium-small scheme. A very large 
fraction of sites, $ \sim (1-I_f)$, 
have small distortions since the corresponding
$g=g_1$. The interfacial sites, all with $g=g_2$, 
have the $I_f$ fraction distributed as shown. 
These $x_i^0$, which are already proportional to the
respective $g_i$, generate the $T=0$ scattering 
potential $V^i_0 = g_i x_i^0$ that decides the 
residual resistivity. We show the distribution
$P(V_0,f) = (1/N) \sum_i \delta(V_0 - g_i x_i^0)$ 
in the Supplement.

Fig.4(a) also marks out three `typical' sites, 1, 2, 3,
where we want to probe thermal fluctuations. We
examine the time series $x_i(t)$ at these sites
at different temperatures and construct the
distribution of fluctuations: $P_i(\delta x)
= (1/N_{\tau}) \sum_{\tau} \delta(\delta x
- (x_i - x_i^0))$, Fig.4(c),
where $N_{\tau}$ is the number of thermal samples.
The thermal scattering would arise from the
`disorder' 
$V_{th}^i = g_i^2 (\Delta x_i)^2$, where 
$(\Delta x_i)^2$ is the variance of  $P_i(\delta x)$
defined earlier. Note that at site 1 - a medium 
density site
- the width of  $P_i(\delta x)$ is more than twice 
the value at sites 1 and 3. 

Fig.4(d) shows that at all the sites the width 
$(\Delta x_i)^2 \propto T$.
The nature of thermal fluctuations is consistent with
a local density dependent phonon stiffness $K_i$
which is close to the bare value for $n_i \sim 0, 1$
and is significantly smaller for $n_i \sim 0.6$.
In the Supplement we show a simple perturbative
calculation, using $P(V_0,f)$ and the thermal
fluctuation $\Delta x_i(T)$, to explain 
the $\rho_0$ and the linear $T$ slope.

On the whole, the residual
resistivity arises primarily from the large
$x_i^0$ sites at the interface, while
the thermal scattering is dominated by the
medium density sites at the interface.

{\it Discussion:}
(i)~Effect of electronic parameter variation:
The $g_1$ value used in our calculation 
should correspond to bulk Au in the
3D Holstein model. We found  
the 3D value to be $\sim 0.3$ (see Supplement).
Given the smaller coupling window for
metallicity in 2D (limited by polaron formation) 
we used the slightly lower value, $g_1 = 0.2$. 
The $g_2$ value is limited by the polaronic instability
which occurs around $g \sim 2$ in 2D. We have explored
$g_2 = 1.4, 1.6$ and find a rapid variation,
faster than $g^2_2$, in the linear $T$ coefficient.
Our $g_2$ value is mainly to  demonstrate a 
qualitative effect. 
We have also checked the effect of 
a higher site potential on Ag sites so that 
there is a 
higher electron density in the Au region. It does
not make much difference as long as $g_1$ and $g_2$
are kept as before. The results are in the Supplement.

(ii)~The possible origin of large interfacial EP
coupling: We have made no suggestions on the physical
mechanism that could generate a large $g_2$, and have
only addressed the consequences of such a coupling.
Addressing this requires modeling the
local chemistry of the Ag-Au system, including 
the local potentials involved, and possible
Ag to Au charge transfer. This has been
attempted \cite{arind2}, at the level of a 
model Hamiltonian,
in the experimental paper itself. Their 
result, at small charge transfer $\delta n$, 
extrapolated to realistic $\delta n$, suggests
that an EP coupling consistent with resistivity
amplification might be achievable. This, however,
is an averaged, homogeneous, 
conception of the EP coupling, while we find
that even the interface is a spatially 
differentiated object. A meaningful
exploration of nanohybrids would involve
DFT calculations to set local parameters
and then many body schemes for transport and
collective effects.

(iii)~Possible superconductivity:
since there is an enhanced, albeit inhomogeneous, EP
coupling in the system, larger than what obtains in
superconducting elements like Pb, Hg, or compounds
like Nb$_3$Sn \cite{allen-epc}, 
one may ask whether this could
lead to phonon induced superconductivity. Our treatment
of phonons is classical, but once a $T=0$ distortion
field $x_i^0$ is obtained one can construct a harmonic
quantum theory of the phonons. The associated
phonon propagator $D_{ij}(\omega)$ could in principle
mediate pairing but the strong disorder in the 
background needs to be considered in any theory
of superconductivity. In addition if there are
Coulombic effects in the real nanohybrid, due to
charge transfer, that too needs to be built in.
A periodic nanohybrid configuration may be a 
better candidate for superconductivity.

{\it Conclusions:}
We constructed a electron-phonon model to address the
huge resistivity amplification
recently seen in Au-Ag nanohybrids. After generating
cluster motifs appropriate to volume fractions
$f$ of Ag in Au, we defined a model that has low
electron-phonon coupling in the `bulk' and a large
coupling on interfacial sites.
We used an exact diagonalisation based Langevin
scheme to handle the electron-phonon coupling and
thermal fluctuations.
For our choice of bulk and interfacial 
coupling we obtained a thermal
resistivity amplification that is consistent
with experiments in both it's volume fraction
dependence and overall magnitude.
More importantly we 
find that while the interface is the crucial 
player in electron 
scattering, the dominant contributions 
to residual resistivity and thermal resistivity
come from different spatial regions of the 
interface. Combined with first principles
estimate of a few electronic parameters in
the nanohybrid, our approach is capable of
predicting the novel transport features
and possible superconducting instabilities
in these systems.

{\it Acknowledgments:} We acknowledge use of the HPC clusters at HRI.
PM thanks Arindam Ghosh for sharing and discussing the experimental
results.


\clearpage

\newpage

\onecolumngrid
\setcounter{figure}{0}

\begin{center}
\fontsize{16pt}{20pt}\selectfont
\textbf{Supplementary to `Enormous enhancement of resistivity in 
\\ nanostructured electron-phonon systems'}
\end{center}

\section{I.~Geometry of the Au-Ag nanostructures}

\subsection{Generating cluster motifs}

 Consider a square lattice 
with $L \times L$ sites of gold
 atoms (Au). Silver (Ag) atoms replace some of the gold in 
this matrix.  
The presence of gold or silver at a site is represented
  by $\eta_i$, a binary field: $\eta_i = 0$ for gold and $\eta_i = 1$ 
  for silver. In the clean limit 
  $\eta_i = 0$ throughout the lattice. For a finite fraction $f$, we 
  randomly select $\zeta$ sites (located at $\vec{R}_{\zeta_i}$) and 
  create a circular region around each of them 
  with radius $2$ (in units of lattice spacing). 
Within each of these regions, the field 
  $\eta_i$ is set to 1. The circles may overlap, and the fraction
   $f$ is given by $(1/L^2)\sum_i \eta_i $.
We define the boundary gold atoms surrounding these islands 
as the interface sites. Everywhere else, the value of $g_i=g_1$
while at the interface sites $g_i=g_2$.
In Fig.1 we show three Au-Ag nanostructures each for
three filling fractions $f$ on a $30\times30$ lattice. 

\begin{figure}[h]
    \centering
    ~~~~~~~~~~~\includegraphics[width=12cm,height=14cm]{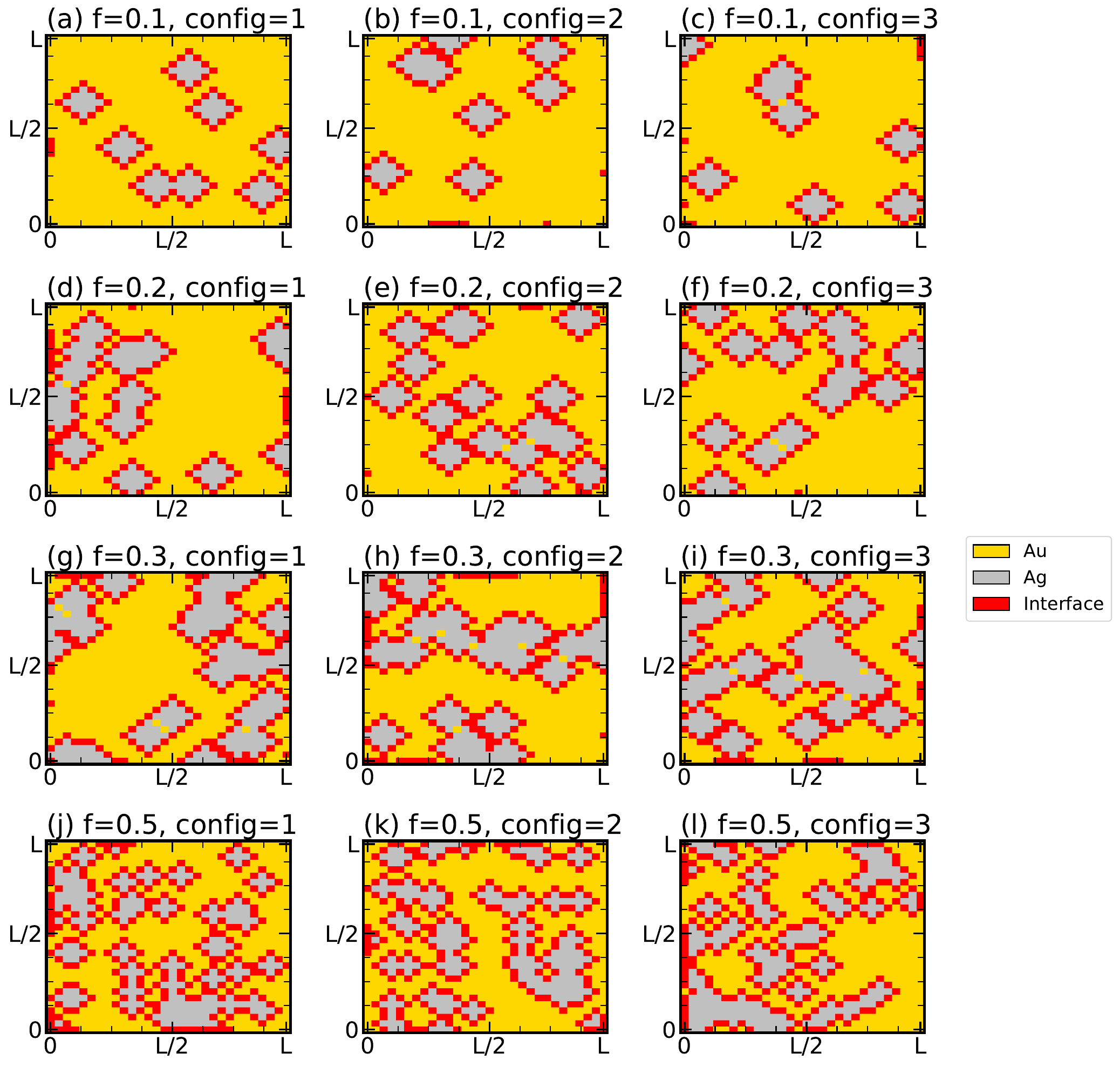}
    \caption{Various configuration maps $\eta_i$ for each filing faction
     $f$ (along the rows) and for different filling fractions $f$ 
     (along the column)}
    \label{fig:figure1}
\end{figure}

\subsection{Interface length}

The fraction of interfacial sites $I_f $ increase with filling fraction 
$f$ upto $f=0.5$. The relationship between $I_f$ and $f$ in the
non-overlapping regime (where the silver droplets are far away from
 each other, i.e. small $f$ regime) is 
$I(f)=\frac{Z}{A}f$
where, $Z$ is the number of interfacial sites in a single cluster 
and $A$ is number of silver sites inside the cluster.
For larger $f$, the droplets starts to overlap and this equation 
overestimates the $I(f)$. This relationship between $I_f$ and $f$,
 averaged over 15 configurations is shown in Fig.2.

\begin{figure}[h]
    \centering
    \includegraphics[width=8cm,height=8cm]{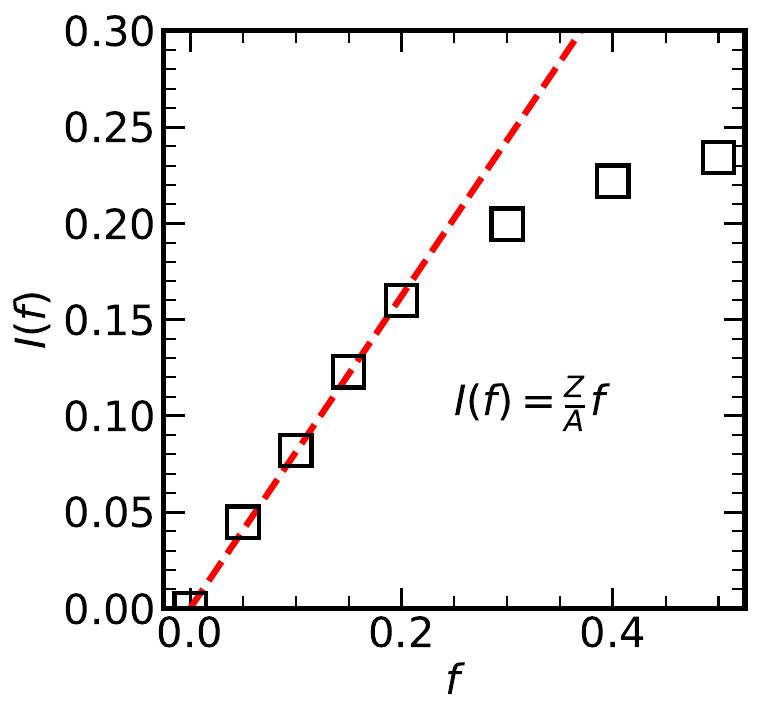}
    \caption{relationship between $I(f)$ and $f$ averaged
     over 15 configurations}
    \label{fig:figure1}
\end{figure}

\vspace{.3cm}

\section{II.~Benchmarking cluster based Langevin 
update with exact diagonalization}

To generate the phonon configurations ${x_i}$, an
 exact diagonalization-based Langevin scheme  
 can be employed. The primary computational challenge lies 
 in calculating the 
force $\langle \partial H/\partial x_i\rangle$, 
 which requires the diagonalization of an $N\times N$ matrix, where 
 $N$ is the total number of sites. This process scales as 
 $\mathcal{O}(N^3)$. The computational cost can be
reduced by considering a smaller cluster centered around the
 site and replacing $\langle \partial H/\partial x_i\rangle$ with 
 $\langle \partial H/\partial x_i\rangle_{\mathcal{C}}$, where the
 force is calculated using the cluster $\mathcal{C}$ surrounding
 the $i$-th site. It is important to note that in this context, the
 term "cluster" refers to a local neighborhood.

In our calculations, we use a cluster consisting of $N_c=13$ sites. 
This cluster includes the site $i$, its 4 nearest neighbors (NN),
 the 4 next-nearest neighbors (NNN), and the 4 axial sites beyond
 them. We define a Hamiltonian $H_{\mathcal{C}}$ on this 
 cluster with open boundary conditions and diagonalize it to 
 compute the force on $x_i$. This force is then used in the
 Langevin equation to generate the equilibrium configurations 
${x_i}$. When calculating electronic properties, we employ
a full diagonalization method.

In Fig.3, we compare the distribution function $P(x)$ obtained 
from both the cluster-based and exact diagonalization-based 
Langevin dynamics in the homogeneous case, where the
 electron-phonon coupling $g$ is uniform across all sites. We 
 also investigate the effect of varying temperature. The results 
 demonstrate that the cluster-based Langevin dynamics closely 
 matches the outcomes of the exact diagonalization-based 
 approach.

\begin{figure}[h]
    \centering
    \includegraphics[width=10cm,height=10cm]{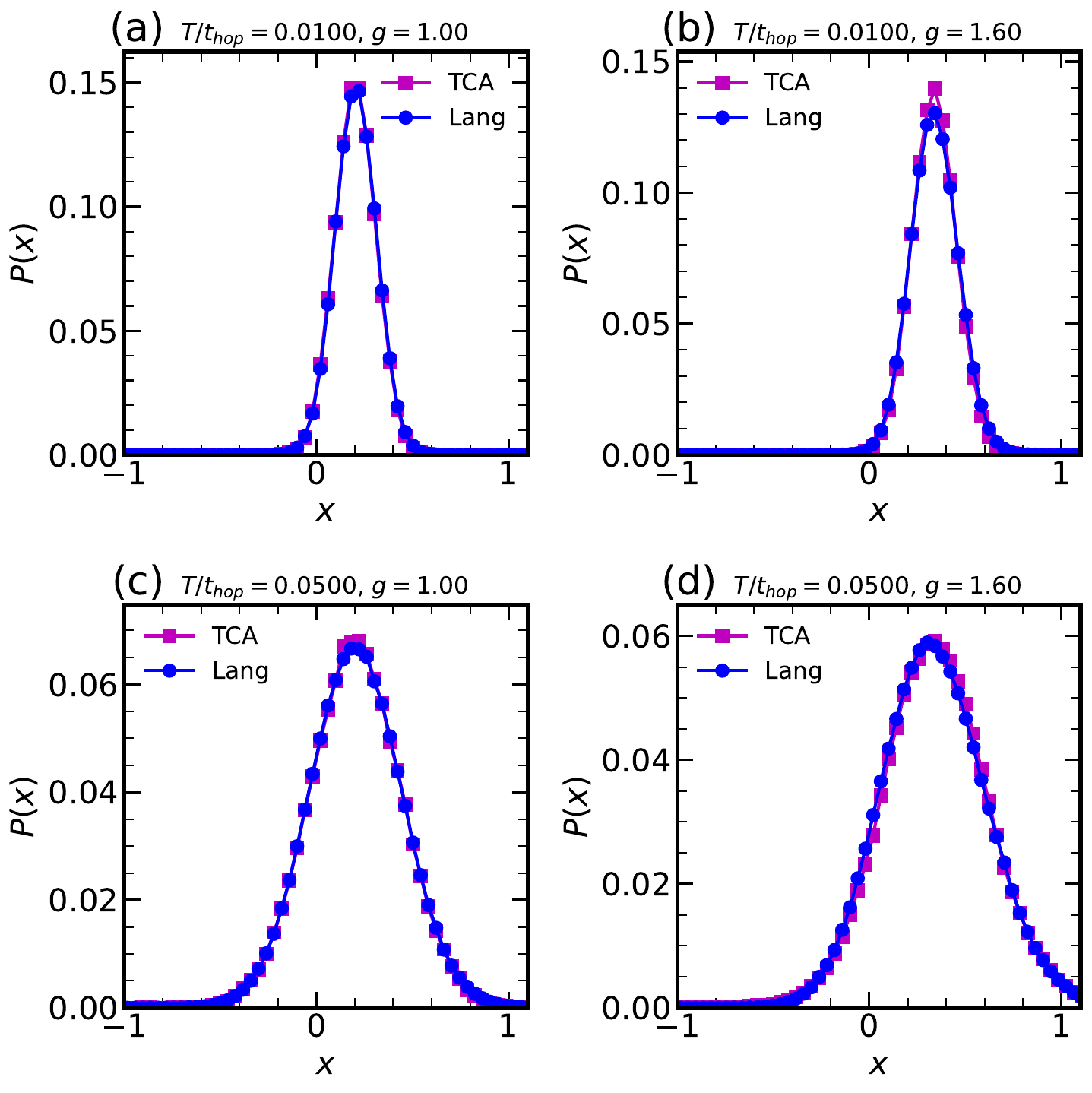}
    \caption{Benchmark of cluster-based $P(x)$ vs $x$ at two 
    temperatures and two values of $g$ with respect to 
    full Langevin calculation on a $20^2$ lattice. The 
    panels correspond to:
(a) $T/t_{\text{hop}}=0.0100$, $g=1.00$;
(b) $T/t_{\text{hop}}=0.0100$, $g=1.60$;
(c) $T/t_{\text{hop}}=0.0500$, $g=1.00$;
(d) $T/t_{\text{hop}}=0.0500$, $g=1.60$.}
\end{figure}

\section{III.~Estimation of $g_1$ for Gold}

To estimate the value of $g_1$, we begin by considering
 the resistivity of gold (Au) at 300 K, which is approximately 
 2 $\mu\Omega\cdot$cm. From perturbative theory, the
  resistivity as a function of temperature is given by:

\[
\rho(T) = \rho_0 \cdot A \cdot g^2 \cdot \frac{T}{t_{hop}}
\]

where $\rho_0 = \frac{\hbar a_0}{\pi e^2}$ is a characteristic
 resistivity, and $A \cdot g^2 \cdot \frac{T}{t_{hop}}$ is 
 dimensionless. The proportionality constant $A$ can be 
 determined from a previous work(\href{https://doi.org/10.1103/PhysRevLett.94.136601}{Phys. Rev. Lett. 94, 136601 (2005)}) on disordered
  Holstein models, where for $g=1$, it was found that $\rho(T)
   = 4 \rho_0$ at $T=0.3t_{hop}$. Thus, $A = \frac{4}{0.3} 
\approx 13$.
For gold, with $a_0 = 4 \, \text{Å}$, $\rho_0 \approx 50 \,
 \mu\Omega\cdot$cm. Substituting into the resistivity 
 equation, we have:

\[
50 \cdot 13 \cdot g_1^2 \cdot \frac{300}{t_{hop}} = 2
\]

Solving for $g_1^2$ gives:

\[
g_1^2 = \frac{t_{hop}}{98000}
\]
Assuming a rough estimate for the bandwidth $t_{hop}
 \approx 10000$ K from band structure, we find:
\[
g_1^2 \approx \frac{1}{10} \quad \text{or} \quad g_1 \approx 0.3
\]
This value of $g_1$ is appropriate for a 3D system, where 
the critical coupling for polaron formation is higher than in
 2D. For comparison, in 2D, $g_1 \approx 0.2$ would be
  reasonable, as we consider limiting $g_2$ to around 1.6. 
\newpage

\section{IV.~Characterising the phonon distortions}

In Fig. 4, we present the density map $n^0_i$ [4(a)-(c)],
 corresponding distributions $P(n)$ [4(d)-(f)], lattice distortion 
 field $x^0_i$ [4(g)-(i)], and its distribution $P(x)$ [4(j)-(l)] at
  $T/t_{hop}=10^{-5}$, along with the time-averaged map 
  $(\delta x_i )^2$ [4(m)-(o)].

\begin{figure}[h]
    \centerline{
    ~~~~~~~~~~~~~~~\includegraphics[width=12cm,height=4cm]{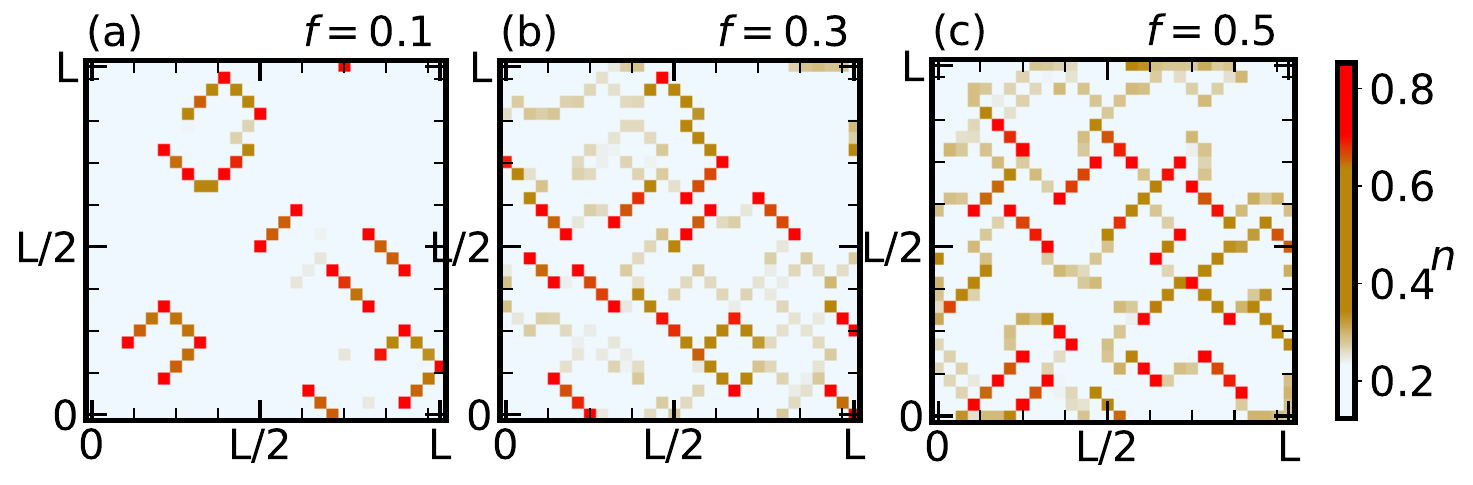} %
    }
    \centerline{
    \includegraphics[width=11cm,height=3cm]{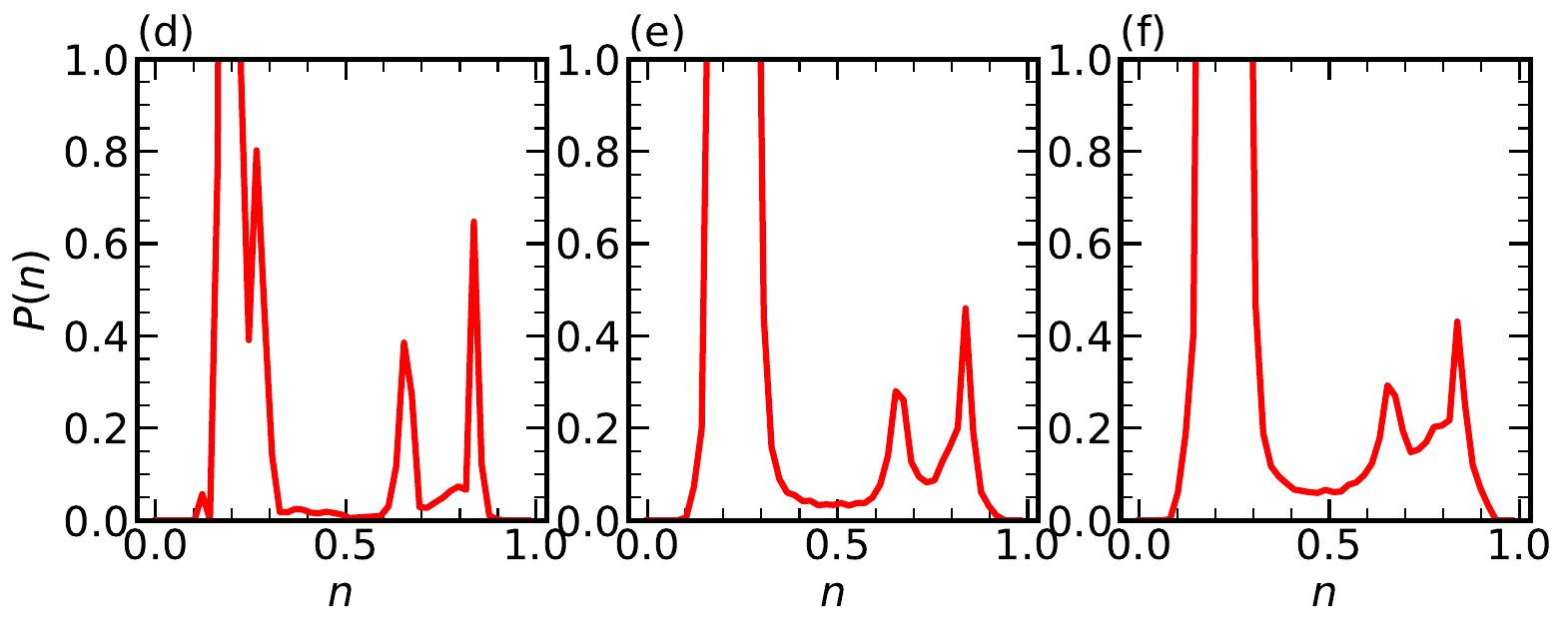} %
    }
    \centerline{
    ~~~~~~~~~~~~~~~\includegraphics[width=12cm,height=4cm]{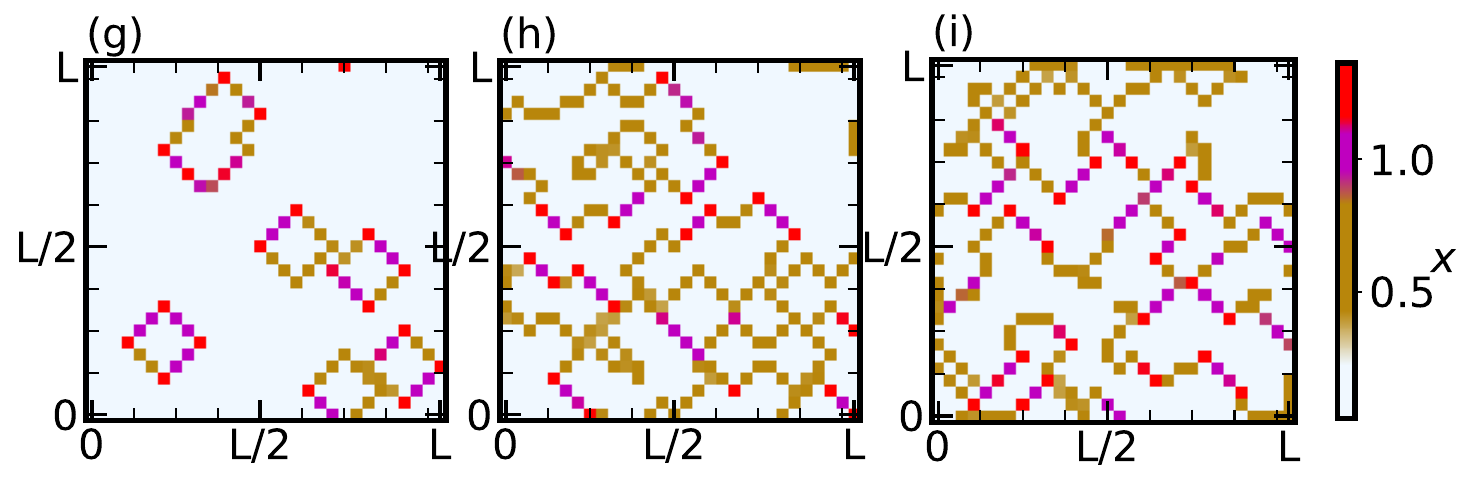} %
    }
    \centerline{
    \includegraphics[width=11cm,height=3cm]{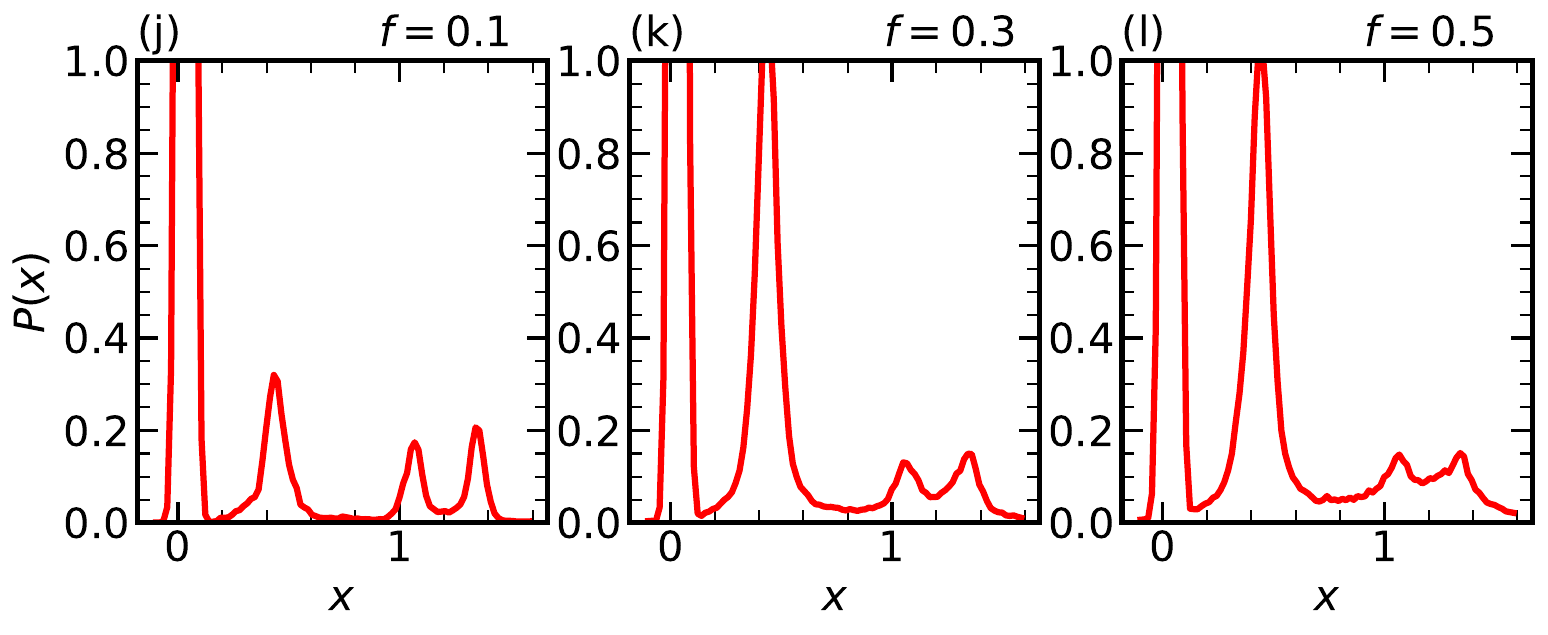} %
    }
    \centerline{
    ~~~~~~~~~~~~~~~\includegraphics[width=12cm,height=4cm]{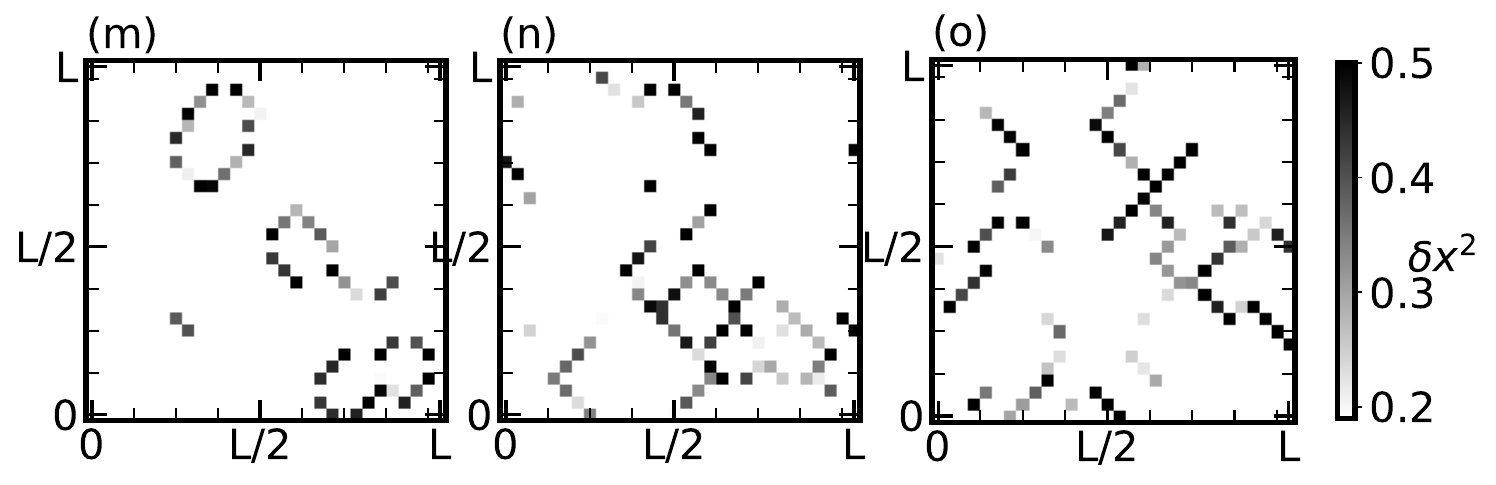} %
    }
    \caption{Density map $n^0_i$ [4(a)-(c)], corresponding 
    distributions $P(n)$ [4(d)-(f)], lattice distortion field $x^0_i$ [
    4(g)-(i)], its distribution $P(x)$ [4(j)-(l)] at $T/t_{hop}=10^{-5}$, 
    and time-averaged map $(\delta x_i)^2$ [4(m)-(o)] at higher $T$.}.
\end{figure}

When we plot the low temperature ($T/t_{hop}=10^{-5}$) 
distribution of the densities on the sites $i$ where 
$i\in \{\text{interface}\}$ we see this distribution function shows
 a distinct three peak structure as shown in Fig. 5(a). 
 These tri-modal structure can be fitted with a three Gaussian 
 distributions of different widths and peak locations. The weight 
 of these three separate peaks are $I_{low}$, $I_{mid}$ and
 $I_{high}$ corresponding to the density values low, medium
 and high. upon changing the $f$, the total number of
 interfacial sites $I(f)$ also changes. To answer how 
 does the $I_{low}$, $I_{mid}$ and $I_{high}$ changes 
 as function of $f$, we plot them in Fig. 5(b).

\begin{figure}[h]
    \centering
    \includegraphics[width=7cm,height=6cm]{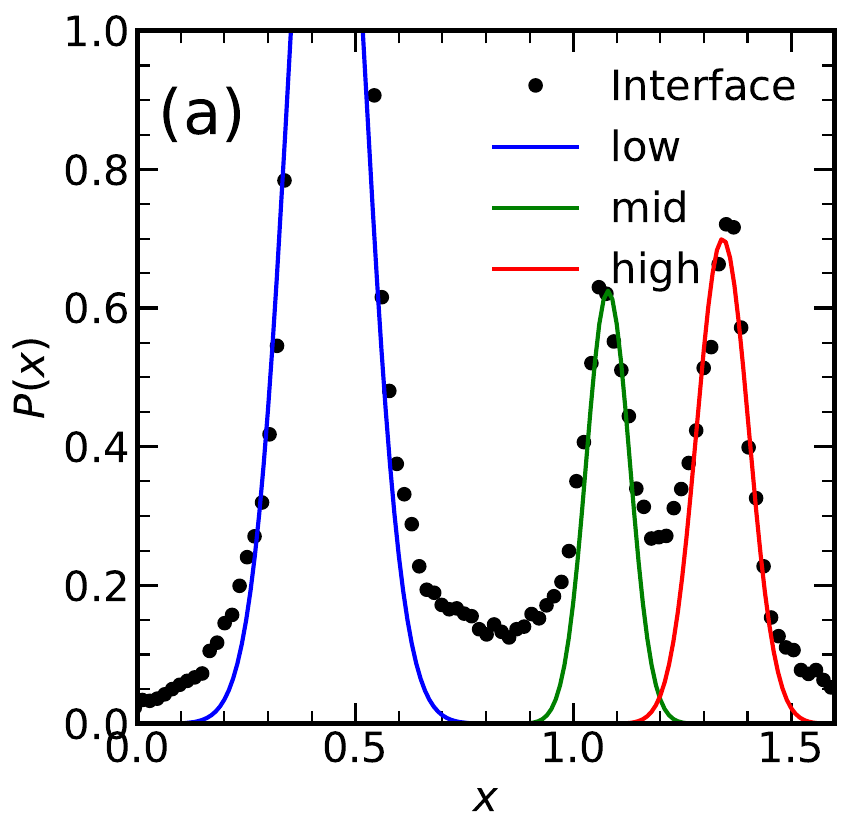}
    \includegraphics[width=7cm,height=6cm]{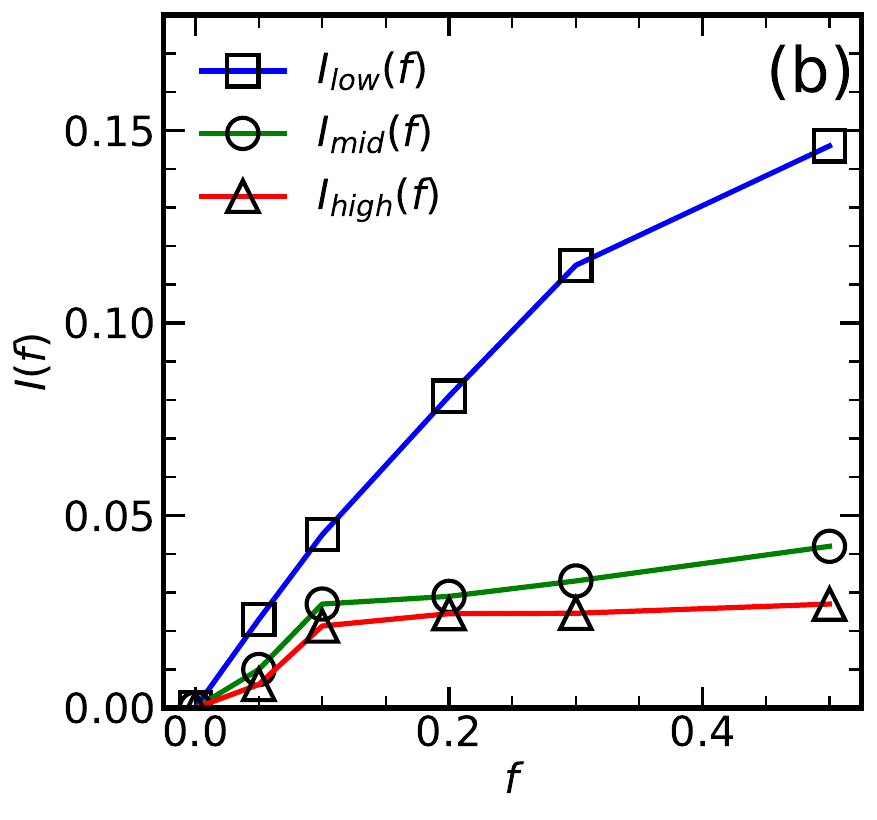} 
    \caption{The distribution of site densities at the interface
     exhibits a distinct tri-modal structure, as shown in Fig. 5(a), 
     which can be fitted with three Gaussian distributions of
     varying widths and peaks. In Fig. 5(b), we plot the evolution
     of the weights $I_{low}$, $I_{mid}$, and $I_{high}$ of
     these peaks as a function of $f$.}
\end{figure}

\section{V. Averaging Conductivity over configurations}

To calculate the conductivity, we assume a separation 
of time scales between charge and phonon dynamics. For an 
instantaneous background $\{x_i\}$, we diagonalize the single 
particle Hamiltonian as described in the main text. This yields 
the eigenfunctions ($f_{i,\epsilon}$) and eigenvalues ($\epsilon_n$). 
The current operator is defined as $j_{mn}^x = 
\langle m|\hat{j}_x|n\rangle$. The conductivity is then computed
as follows:
\begin{equation}
\sigma(\omega) = \sum_{m \neq n}
\frac{|j_{mn}|^2}{\epsilon_m - \epsilon_n}
\delta(\omega - (\epsilon_n - \epsilon_m)) [f_{\epsilon_m}(T) -
f_{\epsilon_n}(T)],
\end{equation}
where $f_{\epsilon}(T)$ denotes the Fermi function.

The D.C. conductivity $\sigma(\omega \rightarrow 0, T)_{\{x_i^{\alpha}\}}$ 
depends on the specific configuration $\{x_i^{\alpha}\}$, where $\alpha$
 denotes different possible realizations of electron-phonon coupling 
 matrix $g_i$ for a given silver-to-gold ratio $f$. These configurations 
 can vary because the local couplings $g^{\alpha}_i$ are generated
  in different ways. In Fig. 5, we plot the D.C. conductivity for various 
  configurations. We then average the D.C. conductivity over all
   configurations (indicated by the black dotted line) to obtain a
    representative value. The resistivity, shown in the inset, is 
    calculated as the inverse of this averaged conductivity. 

\begin{figure}[h]
    \centering
    \includegraphics[width=8cm,height=6cm]{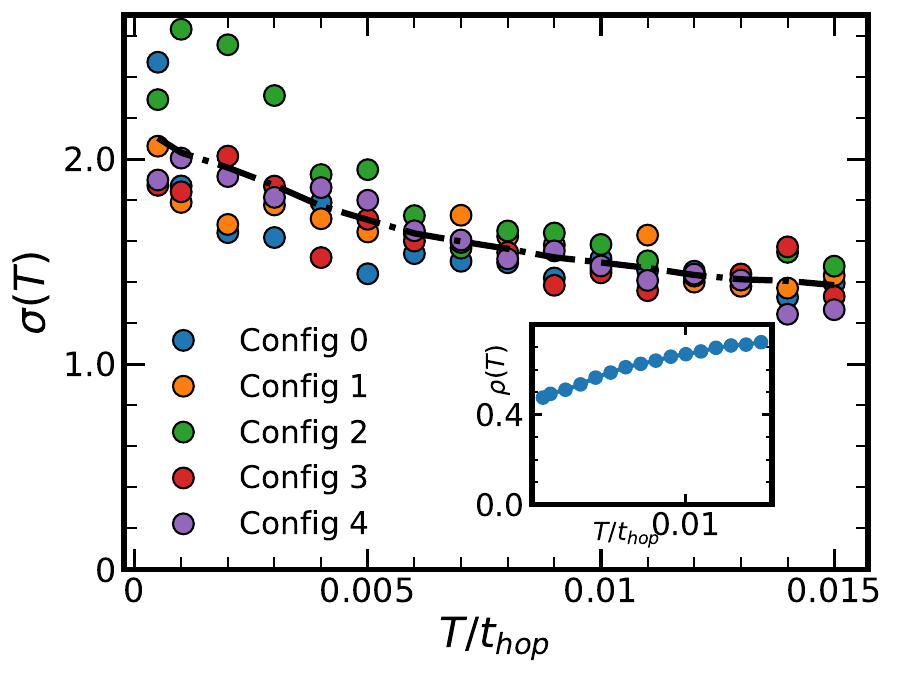} %
    \caption{D.C. conductivity $\sigma$ as a function of 
    temperature $T$ for several configuration corresponding
     to different realization of same $f$ on a $30\times 30$
      lattice. The dotted line shows their average. The plot in
       the inset shows the D.C. resistivity $\rho(T)=1/\sigma(T)$ }.
\end{figure}

\section{VI. Perturbative analysis of resistivity}

\subsection{Calculation for zero-temperature 
resistivity $\rho_0$}

The resistivity $\rho_0$ arising from the 2nd-order 
perturbation theory can be set up if we consider the 
effect of phonons in zero-temperature as a disorder problem. 
The effective disorder potential $V_i$ can be $-g_ix_i$. 
The $x_i$ shows a distinct 4 peak features: 1 peak 
coming from the bulk, valued around $\sim -g_1^2/K$ (a 
small number compared to other cases), 3 other peaks 
arising from the interfacial sites: low (l), medium (m) 
and high (h) values of $x$. The probability distribution 
can be approximated as 
\begin{equation}
    P(x)= (1-I(f))\delta(x-x_{bulk}) + I_{\alpha} \delta(x-x_{\alpha})
\end{equation}
where $I(f)$ is the number of interfacial sites and 
$I_{\alpha}$ denotes the three peaks among those such that 
$I(f)=\sum_{\alpha} I_{\alpha}$ , $\alpha = \{l,m,h\}$. 

\begin{figure}[h]
    \centering
    \includegraphics[width=8cm,height=5cm]{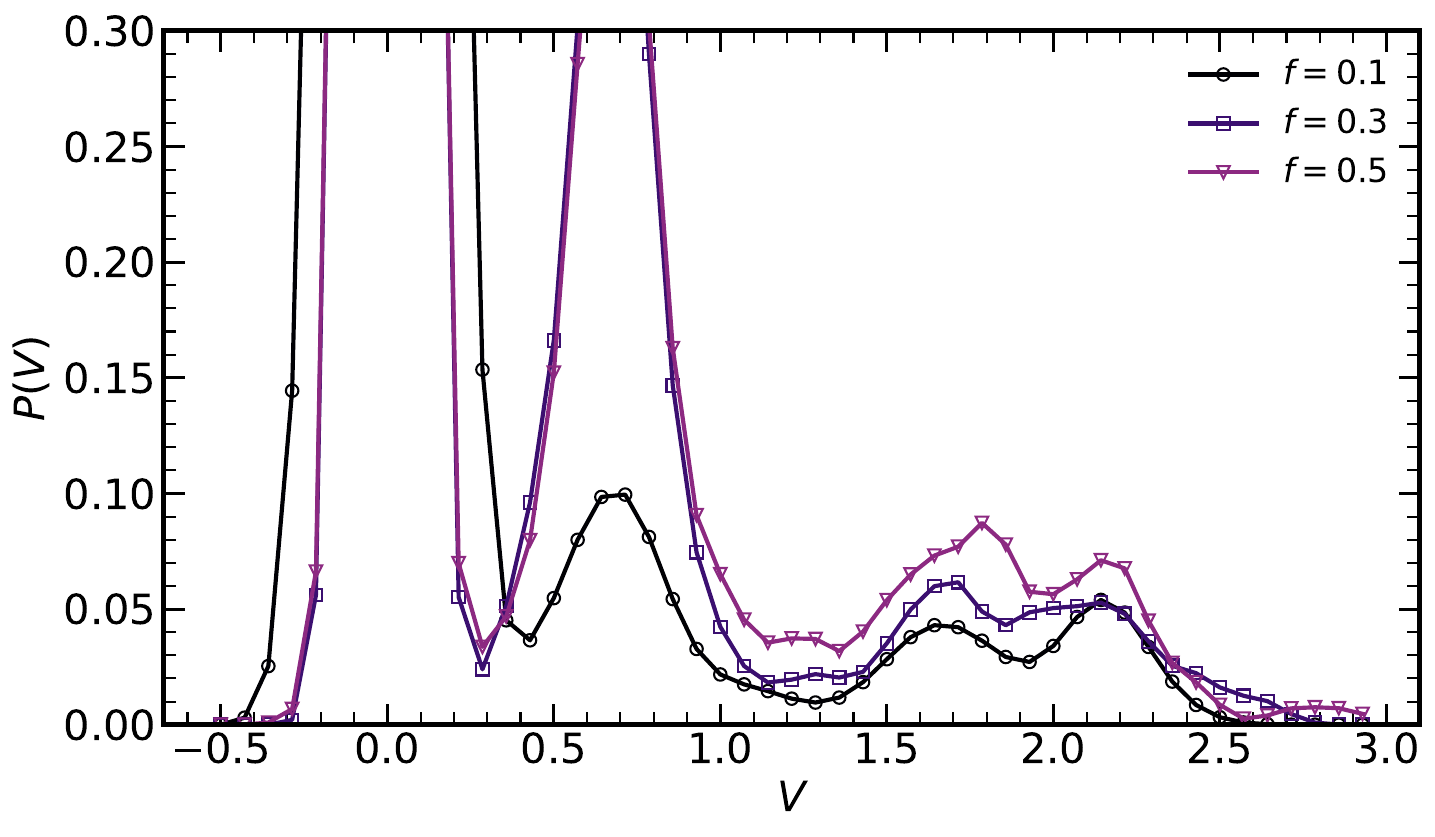} 
    \caption{$P(V)$ for 3 different values of $f$ at
     $T=10^{-5}t_{hop}$,where $V$ is the potential. We approximate this with 
     4 delta functions corresponding to different peaks.}

\end{figure}

The average potential can be written as
\begin{equation}
    \bar{V}=\int V P(V)dV
\end{equation}
$P(V)$ is probability of finding a site with $V$ value 
of potential. In this 4-delta-function approximated 
$x$ distribution we have
\begin{equation}
    \bar{V}=P_{bulk} V_{bulk} + \sum_{\alpha}
     I_{\alpha} V_{\alpha}
\end{equation}
where $V_{bulk}=-g_1 x_{bulk}$ and $V_{\alpha}=-g_2 x_{\alpha}$. 

The resistivity $\rho$ from second order 
perturbation is proportional to 
\begin{equation}
    \rho_0 \propto \langle \Delta V^2\rangle
\end{equation}
And we notice, 
\begin{equation}
     \langle V^2\rangle = \int dV
 V^2 \sum_{\alpha} I_{\alpha} 
\delta(V-V_{\alpha})=\sum_{\alpha} I_{\alpha} V_{\alpha}^2
 \end{equation}
 and
\begin{equation}
    \langle \bar{V} \rangle^2 = \left(\sum_{\alpha} I_{\alpha} 
    \bar{V}_{\alpha}\right)^2 = \sum_{\alpha}
     I_{\alpha}^2 \bar{V}_{\alpha}^2
\end{equation}
Ignoring $g_1x_1^2$ we get,
\begin{equation}
    \rho_0 \propto \langle \Delta V^2\rangle = 
    g_2^2\left[ x_l^2(I_l-I_l^2) + x_m^2(I_m-I_m^2) +
     x_h^2(I_h-I_h^2) \right]
\end{equation}

\subsection{Linear dependence of $\delta\rho$ on $T$}

Now we consider the effect of temperature on resistivity 
by assuming that the distortions $x_i$ can fluctuate around 
their $T=0$ value $x_i^0$. This results in a broader distribution 
around the peaks. This broadening of the distribution $P(\delta x)$
 is shown in Fig.4(c) of the main paper where we plotted data for
  3 sites in \(P(\delta x)\), and we find that the standard deviation 
  of sites corresponding  to $x$ by a factor of almost 6 for the 
  middle-density site ($\alpha=m$), whereas it remains the same
   for the small-density ($\alpha=l$) or large-density sites 
   ($\alpha=h$). This changes the resistivity as
\begin{equation}
    \rho(T) \sim \rho_0 + \delta\rho (T)
\end{equation}
We use this information in \(\delta \rho(T)\) to estimate the
 \(A\) coefficient using this perturbative technique.

In a similar way we now calculate the resistivity. 
$\langle V\rangle^2$ shows no change due to this fluctuations
 but a correction comes from $\langle V^2\rangle \sim g_1^2(1-I(f))(\langle  x_{bulk}^2 \rangle + \langle \delta x_{bulk}^2 \rangle) + \sum_{\alpha} 
 I_{\alpha} g_2^2 (\langle x_{\alpha}^2\rangle +\langle \delta
  x_{\alpha}^2\rangle)$. The extra correction to resistivity it adds:
\begin{equation}
    \delta \rho(T)_f \sim g_1^2(1 - I(f))\langle \delta x_{bulk}^2 \rangle
    + g_2^2(I_l\langle \delta x_l^2 \rangle + I_m \langle \delta
     x_m^2 \rangle + I_h\langle \delta x_h^2 \rangle)
\end{equation}
With $\langle \delta x_m^2 \rangle = 6 \langle \delta x_h^2 
\rangle$ and $\langle \delta x_l^2 \rangle = \langle \delta
 x_h^2 \rangle= \langle \delta
 x_{bulk}^2 \rangle =\delta x^2$, we have
\begin{equation}
    \delta \rho(T)_f \sim 
 g_2^2(I_l + 6I_m + I_h)\delta x^2 + g_1^2(1 - I(f))\langle \delta x^2 \rangle
\end{equation}
The standard deviation $\delta x^2 \propto T = a
 \cdot T$ (Fig 4(d) in main paper), $a$ being a constant. 
\begin{equation}
    \delta \rho(T)_f = a~g_1^2 \langle \delta
     x^2 \rangle + a~ g_2^2 (I_l + 6I_m+I_h)\cdot T
\end{equation}
To eliminate the dependence of $a$ we normalize 
$\delta\rho(T)_f/\delta\rho(T,f=0)$
and the linearity coefficient $A$ comes out to be:
\begin{equation}
    A \sim \frac{g_1^2(1-I(f))+ g_2^2 (I_l + 6I_m+I_h)}{g_1^2} 
\end{equation}

\subsection{Comparing with numerical results}

We estimated $\rho_0$ and $A$ (via $\delta\rho(T)$) 
using perturbative analysis. In Fig.8 we plot these two 
quantities as a function of $f$ against the exact
 numerical values. In 8(a) we plot the resistivity $\rho_0$
  normalized by $\rho_0(f=0.05)$. In Fig.8(b) we plot $A(f)$.
 Both these results matches 
quite well with the numerical exact results. 

\begin{figure}[h]
    \centering
    \includegraphics[width=5.5cm,height=5cm]{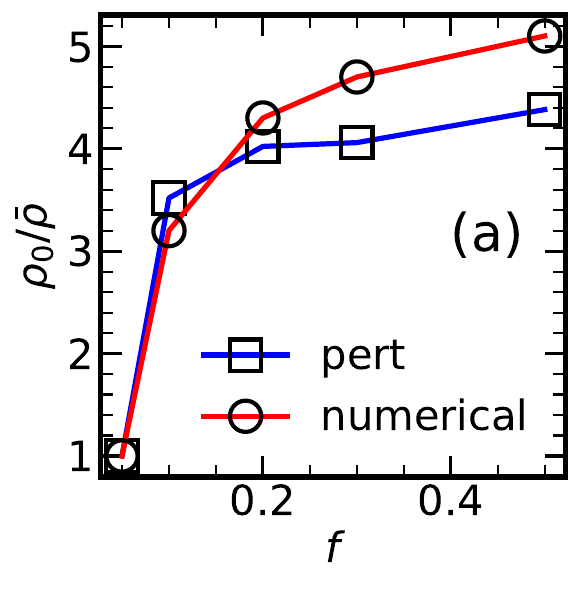} 
    \includegraphics[width=5.5cm,height=5cm]{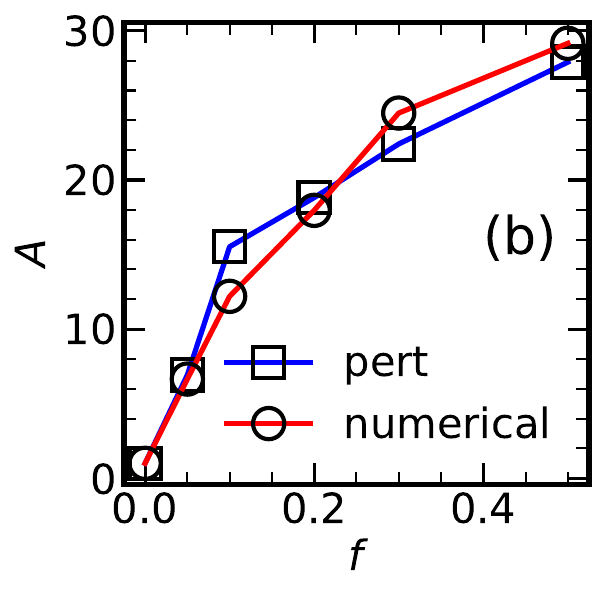} 
    \caption{Comparison of perturbative estimates and exact
     numerical values. Figure (a) shows the resistivity 
     \(\rho_0\) normalized by \(\rho_0(f=0.05)\) as a
      function of \(f\). Figure (b) displays the coefficient \(A(f)\).}

\end{figure}

\section{VII. Effect of parameter variation}

\subsection{variation of $g_2$}

In Fig.9., we plot the resistivity $\rho(f=0.5,T)$ at 
$f=0.5$ for two different $g_2$ values (1.4 and 1.6).
 Notice, the ratio of zero temperature resistivity for
  these two different $g_2$ values is 
  $\rho(T=0,g_2=1.6)/\rho(T=0,g_2=1.4)\sim 7$. The
  linear coefficient $\lambda$ can be extracted 
  ($\rho(T)\sim \rho_0 + \lambda T/t_{hop}$) and 
  $\lambda(g_2=1.6)/\lambda(g_2=1.4)=2.5$. 
\begin{figure}[h]
    \centering
    \includegraphics[width=5.5cm,height=5cm]{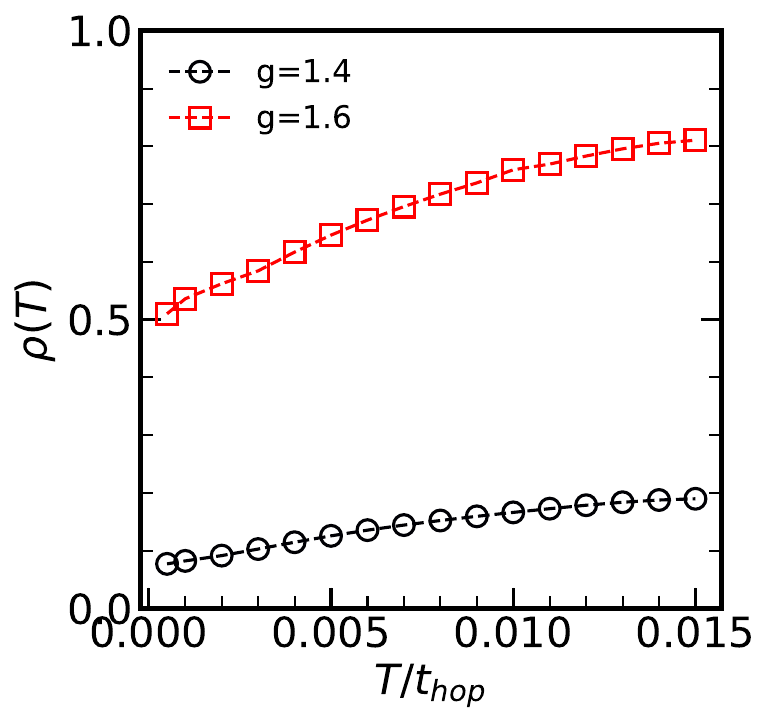}
    \caption{The effect of varying \( g_2 \) values on 
    \(\rho(T)\) is investigated. Calculations are
     performed for \( g_2 = 1.4 \) and \( g_2 = 1.6 \).}
\end{figure}

\subsection{Effect of onsite potential $\mu_i$}

Till now we did not consider the charge transfer between
 Au-Ag. In a simplified picture this would be equivalent to
  introducing an on-site term $-\sum_i\mu_i n_i$ to the 
  Hamiltonian where $\mu_i$ is set to chemical potential 
  $\mu_{Ag}$ ($\mu_{Au}$) for Silver (Gold) sites. We consider
   this in the following calculation and set $\mu_{Au}-
   \mu_{Ag}=0.5t_{hop}$. In Fig.10(a), we plot the density 
   map $n_i$ for one such configuration at $f=0.5$ and 
   $T/t_{hop}=10^{-5}$. This shows the net charge is
    transferred from the Ag sites to Au sites. 
The resistivity as a function of temperature $\rho(T)$ is 
plotted in Fig.10(b) for three different $f$-values which 
shows trend similar to the case with out $mu_i$. 
\begin{figure}[h]
    \centering
    \includegraphics[width=5.5cm,height=5cm]{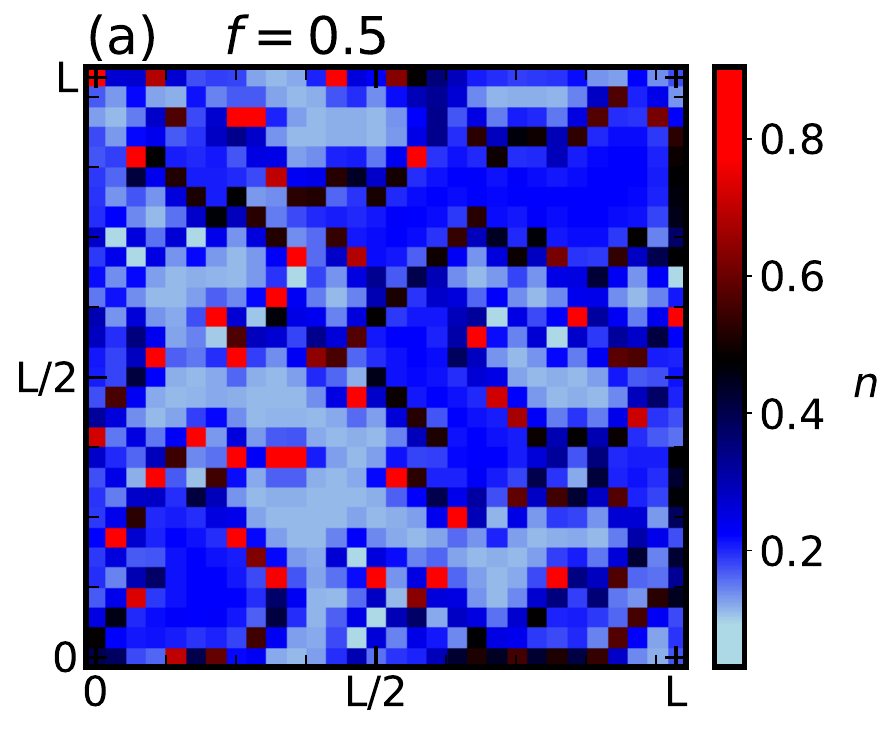} 
    \includegraphics[width=5.5cm,height=5cm]{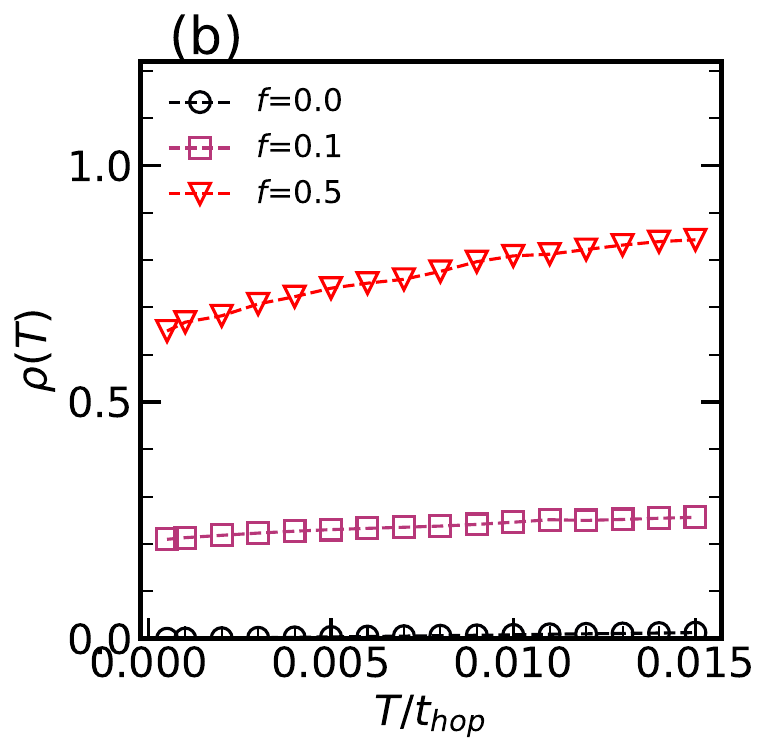} 
    \caption{(a)~Density map \( n_i \) for a configuration
     with charge transfer between Au and Ag sites at
     \( f = 0.5 \) and \( T/t_{hop} = 10^{-5} \). The charge
     transfer is modeled by setting \( \mu_{Au} -
     \mu_{Ag} = 0.5t_{hop} \). (b)~Resistivity as a 
     function of temperature $\rho(T)$ is plotted
     for three different $f$-values.}
\end{figure}


\begin{thebibliography}{0}%
\makeatletter
\providecommand \@ifxundefined [1]{%
 \@ifx{#1\undefined}
}%
\providecommand \@ifnum [1]{%
 \ifnum #1\expandafter \@firstoftwo
 \else \expandafter \@secondoftwo
 \fi
}%
\providecommand \@ifx [1]{%
 \ifx #1\expandafter \@firstoftwo
 \else \expandafter \@secondoftwo
 \fi
}%
\providecommand \natexlab [1]{#1}%
\providecommand \enquote  [1]{``#1''}%
\providecommand \bibnamefont  [1]{#1}%
\providecommand \bibfnamefont [1]{#1}%
\providecommand \citenamefont [1]{#1}%
\providecommand \href@noop [0]{\@secondoftwo}%
\providecommand \href [0]{\begingroup \@sanitize@url \@href}%
\providecommand \@href[1]{\@@startlink{#1}\@@href}%
\providecommand \@@href[1]{\endgroup#1\@@endlink}%
\providecommand \@sanitize@url [0]{\catcode `\\12\catcode `\$12\catcode `\&12\catcode `\#12\catcode `\^12\catcode `\_12\catcode `\%12\relax}%
\providecommand \@@startlink[1]{}%
\providecommand \@@endlink[0]{}%
\providecommand \url  [0]{\begingroup\@sanitize@url \@url }%
\providecommand \@url [1]{\endgroup\@href {#1}{\urlprefix }}%
\providecommand \urlprefix  [0]{URL }%
\providecommand \Eprint [0]{\href }%
\providecommand \doibase [0]{https://doi.org/}%
\providecommand \selectlanguage [0]{\@gobble}%
\providecommand \bibinfo  [0]{\@secondoftwo}%
\providecommand \bibfield  [0]{\@secondoftwo}%
\providecommand \translation [1]{[#1]}%
\providecommand \BibitemOpen [0]{}%
\providecommand \bibitemStop [0]{}%
\providecommand \bibitemNoStop [0]{.\EOS\space}%
\providecommand \EOS [0]{\spacefactor3000\relax}%
\providecommand \BibitemShut  [1]{\csname bibitem#1\endcsname}%
\let\auto@bib@innerbib\@empty
\end{thebibliography}%


\begin{thebibliography}{100} 
\bibitem{ziman} J. M. Ziman, {\it Electrons and phonons},
Oxford University Press (1960).  

\bibitem{allen-transp}  P. B. Allen, Chapter 6,
``Electron transport'' in
{\it Contemporary Concepts of Condensed Matter Science},
Volume 2, 2006, Pages 165-218.

\bibitem{res-gold}
Nguyen Quang Hoc, Bui Duc Tinh and Nguyen Duc Hien, {\it
Influence of temperature and pressure on the electrical
resistivity of gold and copper upto 1350K and 100 GPa},
\href{https://doi.org/10.1016/j.materresbull.2020.110874}{Materials Res Bulletin, 128 (2020) 110874.}

\bibitem{ep-note} In the literature a dimensionless coupling
$\lambda \propto g^2$ is often called the electron-phonon
coupling. We use $g^2$ since $g$ will be the electron-phonon
coupling strength in our Holstein model.

\bibitem{allen-epc} 
P. B. Allen,  
{\it Electron-phonon coupling constants},
in Chapter 7 of {\it Handbook of Superconductivity},
edited by Charles P. Poole, Jr. Academic Press (2000).

\bibitem{arind1}
Tuhin Kumar Maji, Shreya Kumbhakar, Binita Tongbram, T. Phanindra Sai, 
Saurav Islam, Phanibhushan Singha Mahapatra, Anshu Pandey, and Arindam Ghosh,{\it Electrical resistance in a composite of ultra-small silver nanoparticles embedded in gold nanostructures: implications for interface-enabled functionality},
\href{10.1021/acsaelm.3c00379}{ACS Appl. Electron. Mater. 2023, 5, 2893--2901.}

\bibitem{arind2} 
Shreya Kumbhakar, Tuhin Kumar Maji, Binita Tongbram, Shinjan Mandal, Shri Hari Soundararaj, Banashree Debnath, T. Phanindra Sai, Manish Jain, H. R. Krishnamurthy, Anshu Pandey, and Arindam Ghosh, 
{\it  Engineering ultra-strong electron-phonon coupling and 
nonclassical electron transport in crystalline gold with nanoscale interfaces}, 
\href{https://doi.org/10.48550/arXiv.2405.14684 }{arXiv:2405.14684.}



\bibitem{saturation-allen}
P. B.  Allen, in
{\it  Superconductivity in D- and F-Band Metals},
Edited by H. Suhl and M. B. Maple,  Academic Press, 1980; pp 291–304.

\bibitem{saturation-millis}
A. J. Millis, Jun Hu, and S. Das Sarma, {\it Resistivity saturation revisited: results from a dynamical mean field 
theory},
\href{https://doi.org/10.1103/PhysRevLett.82.2354} {Phys. Rev. Lett. 82, 2354 (1999).}

\bibitem{saturation-gun1}
O. Gunnarsson, M. Calandra, J. Han,  {\it  Colloquium: Saturation of electrical resistivity},
\href{https://doi.org/10.1103/RevModPhys.75.1085}{Reviews of Modern Phys. 2003, 75, 1085.}


\bibitem{saturation-gun2}
M. Calandra, O. Gunnarsson,  
{\it  Saturation of electrical resistivity in metals at large 
temperatures}, 
\href{https://doi.org/10.1103/PhysRevLett.87.266601}{Phys. Rev. Lett. 2001, 87, 266601.}

\bibitem{pol-alexand}
For a general reference see, e.g., A. S. Alexandrov and N. F. Mott,
{\it Polarons and Bipolarons}, 
World Scientific, Singapore 1995.

\bibitem{pol-franchini}
Cesare Franchini, Michele Reticcioli, Martin Setvin 
and Ulrike Diebold,
{\it Polarons in materials}, 
\href{https://doi.org/10.1038/s41578-021-00289-w}{Nature Reviews Materials, 6, 560–586 (2021).}

\bibitem{pol-romero}
Aldo H. Romero, David W. Brown, and Katja Lindenberg,
{\it Polaron effective mass, band distortion, and self-trapping in 
the Holstein molecular-crystal model},
\href{https://doi.org/10.1103/PhysRevB.59.13728}{Phys. Rev. B 59, 13728 (1999).}

\bibitem{pol-bonca}
J. Bonca, S. A. Trugman, and I. Batistic, {\it Holstein polaron},
\href{https://doi.org/10.1103/PhysRevB.60.1633}{Phys. Rev. B 60, 1633 (1999).}

\bibitem{pol-ciuchi} 
S. Ciuchi, F. de Pasquale, S. Fratini, and D. Feinberg,
{\it Dynamical mean-field theory of the small polaron},
\href{https://doi.org/10.1103/PhysRevB.56.4494}{Phys. Rev. B 56, 4494 (1997).}

\bibitem{pol-millis}
A. J. Millis, R. Mueller and B. I. Shraiman, 
{\it Fermi-liquid-to-polaron crossover. I. General results},
\href{https://doi.org/10.1103/PhysRevB.54.5389}{Phys. Rev. B, 54 (1996) 5389.}

\bibitem{pol-pm}
B. Poornachandra Sekhar, Sanjeev Kumar and Pinaki Majumdar,
{\it The many-electron ground state of the adiabatic 
Holstein model in two and three dimensions},
\href{https://doi.org/10.1209/epl/i2004-10226-2}{Europhys. Lett., 68 (4), 564 (2004).}

\bibitem{ep-dis-a15-1}
L. R. Testardi, J. M. Poate, and H. J. Levinstein,
{\it Anomalous electrical resistivity and defects in $A-15$ compounds},
\href{https://doi.org/10.1103/PhysRevB.15.2570}{Phys. Rev. B 15, 2570 (1977).}

\bibitem{ep-dis-a15-2}
H. Wiesmann, M. Gurvitch, A. K. Ghosh, H. Lutz, O. F. Kammerer, and Myron Strongin,
{\it Estimate of density-of-states changes with disorder in $A-15$ superconductors},
\href{https://doi.org/10.1103/PhysRevB.17.122}{Phys. Rev. B 17, 122 (1978).}

\bibitem{ep-dis-emin}
David Emin and M.-N. Bussac, 
{\it Disorder-induced small-polaron formation},
\href{https://doi.org/10.1103/PhysRevB.49.14290}{Phys. Rev. B 49, 14290 (1994)}


\bibitem{ep-dis-pm}
Sanjeev Kumar and Pinaki Majumdar,
{\it Singular effect of disorder on electronic transport in 
strongly coupled electron-phonon systems},
\href{https://doi.org/10.1103/PhysRevLett.94.136601}{Phys. Rev. Lett. {\bf 94}, 136601 (2005)}


\bibitem{ep-dis-ciuchi}
S. Ciuchi, D. DiSante, V. Dobrosavljevic, S. Fratini,
{\it  The origin of Mooij correlations in disordered metals}
\href{https://doi.org/10.1038/s41535-018-0119-y}{npj Quantum Mater. 2018, 3, 44.}

\bibitem{lang-moz}
D. Mozyrsky, M. B. Hastings, and I. Martin,
{\it Intermittent polaron dynamics: Born-Oppenheimer 
approximation out of equilibrium},
\href{https://doi.org/10.1103/PhysRevB.73.035104}{Phys. Rev. B 73, 035104 (2006).}

\bibitem{lang-lu}
Jing-Tao Lü, Mads Brandbyge, Per Hedegard, Tchavdar N.
Todorov, and Daniel Dundas,
{\it Current-induced atomic dynamics, instabilities, 
and Raman signals: Quasiclassical Langevin equation approach},
\href{https://doi.org/10.1103/PhysRevB.85.245444}{Phys. Rev. B 85, 245444 (2012).}


\bibitem{lang-pm1}
Sauri Bhattacharyya, Sankha Subhra Bakshi, Samrat Kadge, and Pinaki Majumdar,
{\it Langevin approach to lattice dynamics in a charge-ordered polaronic system},
\href{https://doi.org/10.1103/PhysRevB.99.165150}{Phys. Rev. B 99, 165150 (2019).}


\bibitem{lang-pm2}
Sauri Bhattacharyya, Sankha Subhra Bakshi, 
Saurabh Pradhan, and Pinaki Majumdar,
{\it Strongly anharmonic collective modes in a coupled 
electron-phonon-spin problem},
\href{https://doi.org/10.1103/PhysRevB.101.125130}{Phys. Rev. B 101, 125130 (2020).}


\end{thebibliography}
\end{document}